\documentclass[journal,onecolumn,letterpaper,draftclsnofoot,12pt]{IEEEtran}

\usepackage{amsmath,amsfonts,amssymb,amsthm,mathtools,accents}
\usepackage{algpseudocode,algorithm}
\usepackage{booktabs,enumerate}
\usepackage{cite}

\usepackage{graphicx}
\graphicspath{{figures/}}
\usepackage{epstopdf}

\usepackage{xcolor}
\usepackage{txfonts}  

\DeclareMathOperator*{\IndFn}{I}
\newcommand{\appropto}{\mathrel{\vcenter{
  \offinterlineskip\halign{\hfil$##$\cr
    \propto\cr\noalign{\kern2pt}\sim\cr\noalign{\kern-2pt}}}}}

\newcommand{\vb}[1]{\mathbf{#1}}

\newcommand{\mfg}{\mathsf{m}}
\newcommand{\nfg}{\mathsf{n}}

\newcommand{\overaccent}[2]{\ensuremath{\accentset{#1}{#2}}}

\newcommand{ \paraml }[1]{ \overaccent{\leftarrow}{#1} }
\newcommand{ \paramr }[1]{ \overaccent{\rightarrow}{#1} }
\newcommand{ \paramL }[1]{ \overaccent{\Leftarrow}{#1} }
\newcommand{ \paramR }[1]{ \overaccent{\Rightarrow}{#1} }

\newcommand{\Tra}{\mathrm{T}}
\newcommand{\Hem}{\mathrm{H}}

\theoremstyle{remark}

\theoremstyle{definition}

\title{Probabilistic Receiver Architecture Combining BP, MF, and EP for Multi-Signal Detection} 
\author{Daniel~J.~Jakubisin,~\IEEEmembership{Student~Member,~IEEE},
	R.~Michael~Buehrer,~\IEEEmembership{Senior~Member,~IEEE}, \\ and
	Claudio~R.~C.~M.~da~Silva,~\IEEEmembership{Senior~Member,~IEEE}%
	\thanks{
		Part of this work has been submitted to the 2016 IEEE Global Commun. Conf.
		
	  D.~J.~Jakubisin and R.~M.~Buehrer are with Wireless@Virginia  Tech,  
		Bradley Dept. of  Electrical and Computer Engineering, 
		Virginia Polytechnic Institute and State University,
		Blacksburg, VA 24061 USA
		(e-mail: djj@vt.edu; buehrer@vt.edu).
		
		C.~R.~C.~M.~da~Silva is with Intel Corporation, 
		Santa Clara, CA 95054 USA 
		(e-mail: claudio.silva@ieee.org).}}

\begin{document}

\maketitle

\begin{abstract}
Receiver algorithms which combine belief propagation (BP) with the mean field (MF) approximation are well-suited for inference of both continuous and discrete random variables.  In wireless scenarios involving detection of multiple signals, the standard construction of the combined BP-MF framework includes the equalization or multi-user detection functions within the MF subgraph.  In this paper, we show that the MF approximation is not particularly effective for multi-signal detection.  We develop a new factor graph construction for application of the BP-MF framework to problems involving the detection of multiple signals.  We then develop a low-complexity variant to the proposed construction in which Gaussian BP is applied to the equalization factors.  In this case, the factor graph of the joint probability distribution is divided into three subgraphs: (i) a MF subgraph comprised of the observation factors and channel estimation, (ii) a Gaussian BP subgraph which is applied to multi-signal detection, and (iii) a discrete BP subgraph which is applied to demodulation and decoding.  Expectation propagation is used to approximate discrete distributions with a Gaussian distribution and links the discrete BP and Gaussian BP subgraphs.  The result is a probabilistic receiver architecture with strong theoretical justification which can be applied to multi-signal detection.
\end{abstract}

\begin{IEEEkeywords}
	Belief propagation, mean field approximation, expectation propagation, iterative receivers, 
	parameter estimation, MIMO, multi-user detection, interference.
\end{IEEEkeywords}

\section{Introduction} \label{sec:intro}
\IEEEPARstart{B}{elief} propagation (BP)---also known as the sum-product 
algorithm---is an effective inference tool for many of the tasks performed by a communications receiver.
Notable examples include decoding, demodulation, 
and multi-user detection~\cite{Kschischang2001,Worthen2001,Wymeersch2007}.
In the Bayesian framework, the data (e.g., symbols) and the parameters 
(e.g., channel coefficients) of the received signal
are modeled as random variables.  This results in
a non-linear observation model which, along with the fact that the model 
variables are both continuous and discrete, makes exact application of BP 
infeasible.
To handle the non-linear observation model, some of the work in the literature
has approximated the BP messages by making Gaussian assumptions
on the message distributions where needed~\cite{Wo2012,Knievel2012}.
However, a more common approach has been to work with parameter estimates
(rather than distributions) computed from ``soft'' symbol estimates. 
The expectation maximization (EM) algorithm was shown to provide theoretical 
justification for this approach where soft symbols estimates take the form of
posterior expectations~\cite{Noels2005,Herzet2007b,Jakubisin2014wcnc,Jakubisin2015tcom}. 
The trade-off with these approaches is the loss of probabilistic information. 
While the estimates (i.e., the messages) are computed taking into account the underlying probability 
distributions, they do not convey probabilistic 
information (e.g., our confidence in the estimate). 



Variational message passing based on the MF approximation 
is another viable alternative to BP for estimation of continuous variables and 
for handling the non-linear observation model~\cite{Wainwright2008,Riegler2013}. 
In contrast to EM, messages computed according to the MF 
approximation \emph{do} convey probabilistic information. 
In fact, EM is a special case of the MF approximation where the messages
are given by Dirac delta functions~\cite{Hu2008}. 
A disadvantage of the MF approximation is that it is not suited for demodulation and decoding tasks where the factor nodes contain hard 
constraints. 
Recognizing that BP and the MF approximation have complementary strengths,
receiver algorithms have been developed which combine these 
algorithms~\cite{Hu2008,Kirkelund2010,Manchon2011}.
Riegler \emph{et al}. provided a theoretical justification for the combined
BP-MF message passing framework in~\cite{Riegler2013}.
The justification is based on the construction of a region-based free
energy approximation analogous to that given for BP by Yedidia \emph{et al}.
in~\cite{Yedidia2005}.
Specifically, in~\cite{Riegler2013} it is shown that fixed points of 
the combined message passing algorithm correspond to stationary points 
of a constrained region-based free energy approximation.  
A notable result from Riegler's work is that the BP-MF framework
provides a consistent rule for when to 
pass \emph{a posteriori} probabilities (posterior beliefs) and when 
to pass extrinsic messages based on the constructed model.

The combined message passing algorithm is particularly applicable to 
communications receivers.  
In this setting, BP
is a generalization of iterative decoding of error correction codes 
and the MF approximation is applicable to the estimation of parameters
such as coefficients of a wireless multipath channel.
The combined message passing algorithm is demonstrated 
in~\cite{Riegler2013} for channel estimation and equalization in an 
orthogonal frequency-division multiplexing (OFDM) system with 
demodulation and decoding.  Subsequent works have applied the BP-MF framework
to various channel estimation 
scenarios~\cite{Senst2011bpmf,Senst2012,Badiu2012bpmf,Badiu2013,Hansen2015arxiv,Barbu2016,Yuan2016arxiv,Zhang2016arxiv}.
Combined BP-MF message passing has also been applied to multi-user code-division multiple access (CDMA)~\cite{Hu2008}, multiple-input multiple-output (MIMO) systems~\cite{Manchon2009mimo,Kirkelund2010,Manchon2011,Badiu2012_coop},
co-channel interference~\cite{Manchon2009cochannel}, and frequency-domain 
equalization~\cite{Zhang2015}.\footnote{Some of these works pre-date~\cite{Riegler2013}
and have used the terms variational message passing / sum-product algorithm (VMP-SP) 
and divergence minimization (DM) to refer to algorithms that 
similarly combine BP and the MF approximation.}
Joint channel estimation is included in the majority of these works as well~\cite{Hu2008,Manchon2009mimo,Manchon2009cochannel,Kirkelund2010,Manchon2011,Badiu2012_coop}. 
In scenarios involving detection of multiple signals,
the standard application of the BP-MF framework is to include the 
equalization task within the MF subgraph.\footnote{We use the term equalization to refer to the un-doing of both multi-stream and multi-user interference.}
As we will show in this paper, the MF approximation is 
a poor choice for detection tasks involving signal separation 
(i.e., signal models with interference-corrupted observations).  
In~\cite{Manchon2011}, the BP-MF framework was applied to 
MIMO-OFDM where multi-stream equalization is within the MF subgraph,
but the algorithm relies upon a generous initialization point.

In this paper, we develop a receiver architecture for multi-signal detection
based on the BP-MF framework.  We show how factorization of the joint 
distribution into distinct observation factors and equalization factors greatly 
improves the detection capability of the structure.
This is because the equalization function of the receiver is now chosen to be within the BP subgraph.  
A consequence of this factorization is that the computational complexity
of BP-based equalization is exponential in the number of 
arriving signals components and the modulation order.
This problem may be circumvented by approximating the domain of the symbol
variables to be continuous random variables and the messages returned from 
BP-based demodulation and decoding as Gaussian distributions.
These two approximations lead to Gaussian BP for the equalization factors
which has a complexity independent of the modulation order and which is
polynomial with respect to the number of interfering signal components.
Thus, in the proposed receiver architecture the factor graph is divided into three subgraphs:
\begin{itemize}
	\item MF subgraph: applied to the observation factors and parameter estimation. 
	The MF algorithm can also serve as a link to BP based estimation of parameters.  
	\item Gaussian BP: when the complexity of discrete BP is too high, Gaussian BP is applied to multi-signal detection. The symbol variables are treated as continuous random variables.
	\item Discrete BP: the sum-product algorithm is applied to demodulation and decoding where the factor functions have hard constraints and the variables are discrete. 
\end{itemize}

The question that remains is how to approximate the discrete 
distributions passed from the symbol variables to the equalization factors
with Gaussian distributions.
A common approach in the technical literature is to match the mean and variance 
of the approximating Gaussian distribution with those of the discrete extrinsic 
or posterior distribution~\cite{Witzke2002,Ping2006,Rossi2008,Jiang2009,Jakubisin2015cochannel,Jakubisin2016cochannel}. 
However, expectation propagation (EP)~\cite{Minka2001expectationpropagation,Hu2006}
provides a theoretically justified approach to computing Gaussian approximations
which has been shown to outperform extrinsic
and posterior approximations~\cite{Senst2011,Sun2015}.
Combined BP-EP has also been applied more broadly to OFDM channel estimation~\cite{Wu2014} and 
massive MIMO~\cite{Wu2015message}, but lacks the flexibility 
of the MF approximation (for example, to incorporate estimation of the noise variance~\cite{Yuan2016arxiv}).
In our receiver architecture, EP is used to link the Gaussian and discrete BP subgraphs.
From these developments, we propose a probabilistic message passing
receiver architecture for multi-signal detection which utilizes BP 
(both discrete and Gaussian), MF, and EP.
Very recently a pre-print has appeared which combines  
BP, the MF approximation, and EP for the purpose of joint phase noise 
estimation and equalization of inter-symbol interference~\cite{Wang2016arxiv}.
While~\cite{Wang2016arxiv} bears a conceptual similarity to our work, 
it differs substantially in that it does not handle multi-signal detection 
and assumes knowledge of the channel coefficients.

Our proposed receiver architecture is suitable for multi-signal detection
in a variety of scenarios including co-channel interference (CCI), MIMO, multi-user MIMO, and non-orthogonal multiple access (NOMA).
Reasonable complexity is maintained through the use of the MF approximation
for the observation factors and Gaussian BP for the equalization factors.  
Parameter estimation may be included in the MF subgraph or, using MF as a link
across the non-linear observation model, other algorithms such as BP
or generalized approximate message passing (GAMP)~\cite{Rangan2011} 
may be applied to estimation~\cite{Yuan2016arxiv}.

The contributions of our paper are summarized as follows:
\begin{itemize}
	\item Development of a factor graph construction for applying BP-MF to multi-signal detection
	\item Development of a low-complexity variant of the proposed construction combining
	BP (both Gaussian and discrete), the MF approximation, and EP.
	\item Derivation of a new MF-based time-domain channel estimator for OFDM signals.
\end{itemize}
In presenting these contributions, the paper is organized as follows. 
We begin in Section~\ref{sec:background} by providing background on 
the BP-MF framework and multi-signal system model.
In Section~\ref{sec:construct}, we develop a factor graph construction 
which maintains the benefits of the MF approximation
and BP in the case of the multi-signal model.
In Section~\ref{sec:approx}, we develop a receiver architecture based on MF, 
Gaussian BP, and Discrete BP. Although the architecture makes an approximation
on the domain of the symbol variables, a solid theoretical foundation is maintained
by applying EP to compute messages between the Gaussian and Discrete BP subgraphs.
In Section~\ref{sec:der_mimo}, we apply the developed receiver architecture to a MIMO-OFDM signal
and in Section~\ref{sec:results} numerical results are provided that demonstrate the
performance of the approach.  
Finally, the paper is concluded in Section~\ref{sec:conclusion}.

\subsubsection*{Notation} Column vectors and matrices are denoted by boldface lowercase and 
uppercase letters, respectively. We use $(\cdot)^\Tra$ and $(\cdot)^\Hem$ to denote transpose and 
conjugate transpose, respectively.  
The multivariate complex Gaussian pdf of $\vb{x}$ is denoted by $\mathcal{CN}( \vb{x} ; \mu , \Sigma )$
where $\mu$ is a vector of the means and $\Sigma$ is the covariance matrix.
The size of set $\mathcal{A}$ is denoted by $|\mathcal{A}|$.
The indicator function is denoted $\IndFn(\cdot)$ and returns a value of 1
when the argument is true and 0 otherwise.
Messages passed along the edges of a factor graph are denoted by sans-serif fonts where 
$\mfg_{f_a \rightarrow x}$ denotes the message from factor node $f_a$ to variable node $x$ and 
$\nfg_{x \rightarrow f_a}$ denotes the message from variable node $x$ to factor node $f_a$.


\section{Background} \label{sec:background}

\subsection{BP-MF Framework}
Let $\boldsymbol{\chi} = [\chi_1, \chi_2, \ldots, \chi_K]^\Tra$ be a 
vector of random variables and let $x_i$ represent a possible 
realization of random variable $\chi_i$.
The joint probability distribution 
$p_{\chi_1,\chi_2,\ldots,\chi_K}(x_1,x_2,\ldots,x_K)$ 
is expressed using vector notation as $p_{\boldsymbol{\chi}}(\vb{x})$.
Throughout this work we use $x_i$ to represent both the random variable
and the possible realizations and write the joint distribution
simply as $p(\vb{x})$.
A region-based free energy is defined with respect to the factorization of
the probability distribution.   Consider the following factorization of 
the probability distribution 
\begin{equation}
	p(\vb{x}) = \prod_{a\in \mathcal{A}} f_a(\vb{x}_a),
	\label{eq:factor_function}
\end{equation}
where $\vb{x}_a \triangleq ( x_i | i \in \mathcal{N}(a))^\Tra$ 
with $\mathcal{N}(a)$ denoting variables which appear as arguments 
of factor $a$ (i.e., neighbors in the resulting factor graph).
The factor graph (equivalently, the factors) are partitioned into
a MF subgraph $\mathcal{A}_{\rm MF}$ and a BP subgraph $\mathcal{A}_{\rm BP}$
where $\mathcal{A}_{\rm MF} \cap \mathcal{A}_{\rm BP} = \varnothing$ 
and $\mathcal{A} \triangleq \mathcal{A}_{\rm MF} \cup \mathcal{A}_{\rm BP}$.
The variables associated with each portion of the graph are given by
\[ \mathcal{I}_{\rm MF} \triangleq \bigcup_{a \in \mathcal{A}_{\rm MF}} \mathcal{N}(a) \quad \text{and} \quad \mathcal{I}_{\rm BP} \triangleq \bigcup_{a \in \mathcal{A}_{\rm BP}} \mathcal{N}(a), \]
respectively. 
From this definition, the region based free energy is constructed 
from the following regions \cite{Riegler2013}
\begin{enumerate}
	\item a single MF region containing all factors and variables in the MF portion with a counting number of 1.
	\item large region from the Bethe free energy $R_a \triangleq (\mathcal{N}(a),\{a\})$, with a counting number of 1,
	\item small regions from the Bethe free energy $R_i \triangleq (\{i\},\varnothing)$, with counting number $1-|\mathcal{N}_{\rm BP}(i)| - \IndFn(i \in \mathcal{I}_{\rm MF})$.
\end{enumerate}
The region based free energy is given by~\cite{Riegler2013}
\begin{align}
	F_{\rm BP, MF} = &\sum_{a \in \mathcal{A}_{\rm BP}} \sum_{\vb{x}_a} b_a(\vb{x}_a) \ln \frac{b_a(\vb{x}_a)}{f_a(\vb{x}_a)} \notag \\
		&\quad - \sum_{a \in \mathcal{A}_{\rm MF}} \sum_{\vb{x}_a} \prod_{i \in \mathcal{N}(a)} b_i(x_i) \ln f_a(\vb{x}_a) \notag \\
		&\quad - \sum_{i \in \mathcal{I}} (|\mathcal{N}_{\rm BP}(i)|-1) \sum_{x_i} b_i(x_i) \ln b_i(x_i)
	\label{eq:BPMF_free_energy}
\end{align}
along with constraints for the factorization of the MF portion beliefs, 
normalization constraints, and marginalization constraints as 
detailed in~\cite{Riegler2013}.

The combined BP-MF message passing rules are as follows.
The messages from factor nodes to variable nodes within the MF subgraph
are given by~\cite{Riegler2013}
\begin{equation}
	\mfg^{\rm MF}_{f_a \rightarrow x_i}(x_i) = \exp \left( \sum_{\vb{x}_a \backslash x_i} \prod_{j \in \mathcal{N}(a) \backslash i} n_{x_j \rightarrow f_a}(x_j) \ln f_a(\vb{x}_a) \right)
	\label{eq:mfbp_m_mf}
\end{equation}
for all $a \in \mathcal{A}_{\rm MF},i\in \mathcal{N}(a)$.  
The messages from factor nodes to variable nodes within the BP subgraph 
are given by~\cite{Riegler2013}
\begin{equation}
	\mfg^{\rm BP}_{f_a \rightarrow x_i}(x_i) = \sum_{\vb{x}_a \backslash x_i} f_a(\vb{x}_a) \prod_{j \in \mathcal{N}(a) \backslash i} n_{x_j \rightarrow f_a}(x_j) 
	\label{eq:mfbp_m_bp}
\end{equation}
for all $a \in \mathcal{A}_{\rm BP},i\in \mathcal{N}(a)$. 
Finally, messages passed from variable nodes to factor nodes 
throughout the entire graph are given by~\cite{Riegler2013}
\begin{equation}
	\nfg_{x_i \rightarrow f_a}(x_i) = \prod_{c \in \mathcal{N}_{\rm BP}(i) \backslash a} \mfg^{\rm BP}_{f_c \rightarrow x_i}(x_i) \prod_{c \in \mathcal{N}_{\rm MF}(i)} \mfg^{\rm MF}_{f_c \rightarrow x_i}(x_i) 
	\label{eq:mfbp_n}
\end{equation}
for all $a \in \mathcal{A},i\in \mathcal{N}(a)$ where 
$\mathcal{N}_{\rm BP}(i) = \mathcal{N}(i) \cap \mathcal{A}_{\rm BP}$
and 
$\mathcal{N}_{\rm MF}(i) = \mathcal{N}(i) \cap \mathcal{A}_{\rm MF}$.
In \eqref{eq:mfbp_n}, the messages to factor nodes in the BP subgraph are
extrinsic messages (as denoted by the exclusion of $i$ in the first product).
On the other hand, the messages to factor nodes in the MF subgraph are posterior beliefs.

\subsection{Exemplary Multi-Signal System Model}
Consider the reception of $N$ signals where column vector $\vb{b}_i$ denotes 
the information bits corresponding to the $i$th signal.
The information bits $\vb{b}_i$ are encoded with an error correction
code to produce a vector of coded bits $\vb{c}_i$ and, subsequently, 
modulated using a digital phase-amplitude modulation. 
The resulting complex symbol sequence is denoted by
$\vb{x}_i = \left[ x_i(0), x_i(1), \ldots, x_i(K-1) \right]^\Tra$.
The information bits, coded bits, and symbols for 
all signals are denoted by $\vb{B} = [\vb{b}_1,\ldots, \vb{b}_N]$, 
$\vb{C} = [\vb{c}_1,\ldots, \vb{c}_N]$, and 
$\vb{X} = [\vb{x}_1,\ldots, \vb{x}_N]$, respectively.

\begin{figure}
	\centering
		\includegraphics[width=0.5\columnwidth]{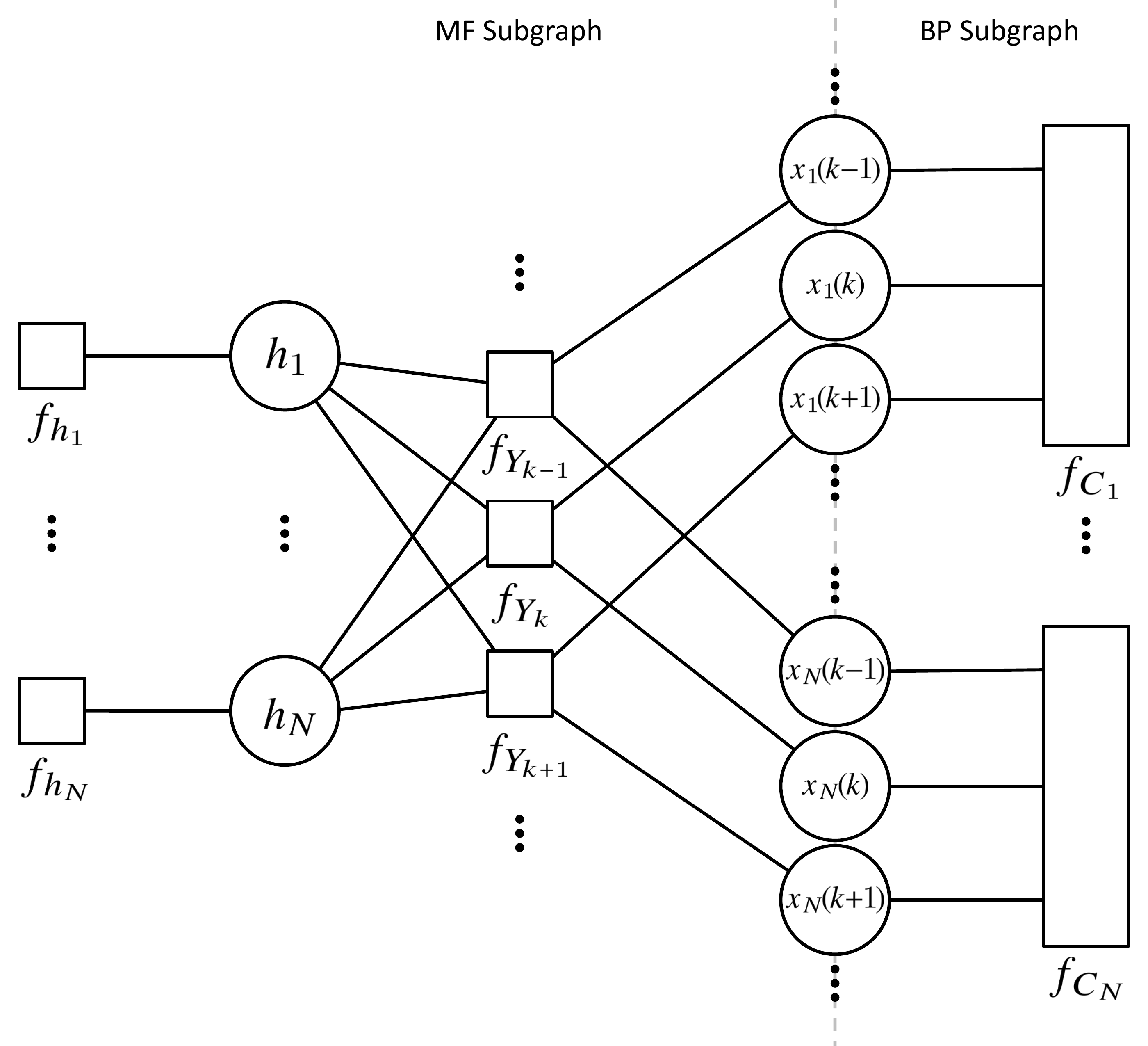}
		\caption{Factor graph of \eqref{eq:joint_example} based on standard construction of BP-MF for multi-signal problems. In the graph, $f_{h_i}$ labels the factor $p(h_i)$, $f_{Y_k}$ labels the factor $p(y(k) | \vb{x}(k),\vb{h})$, and $f_{C_i}$ labels the factors $p(\vb{x}_i|\vb{c}_i)p(\vb{c}_i|\vb{b}_i)p(\vb{b}_i)$.}
		\label{fig:fg_orig}
\end{figure}

The $k$th observation $y_k$ is comprised of $N$ interfering 
signal components $\vb{x}(k) = [x_1(k),\ldots,x_N(k)]^\Tra$ and
white Gaussian noise as given by
\begin{equation}
	y(k) = \sum_{n=1}^N h_n x_n(k) + w(k),
	\label{eq:simple_model}
\end{equation}
where $\vb{h} = [h_1,\ldots,h_N]^\Tra$ are the channel coefficients corresponding 
to the $N$ signals and $w(k)$ are independent identically distributed (iid) circularly-symmetric complex Gaussian random variables with variance $\gamma^{-1}$.
Let the vector of observations be denoted by $\vb{y} = [y(0),\ldots,y(K-1)]^\Tra$ and 
the symbols associated with the $k$th observation be denoted by $\vb{x}(k) = [x_1(k), \ldots, x_N(k)]^\Tra$.
The joint distribution is factored as follows:
\begin{equation}
	p(y,\vb{X},\vb{C},\vb{B},\vb{h}) = \prod_{k=0}^{K-1} \underbrace{p(y(k) | \vb{x}(k),\vb{h})}_{f_{Y_k}} \left( \prod_{i=1}^N \underbrace{p(\vb{x}_i|\vb{c}_i)p(\vb{c}_i|\vb{b}_i)p(\vb{b}_i)}_{f_{C_i}} \right) \left( \prod_{i=1}^N \underbrace{p(h_i)}_{f_{h_i}} \right).
	\label{eq:joint_example}
\end{equation}
The factors $p(\vb{x}_i|\vb{c}_i)$ and $p(\vb{c}_i|\vb{b}_i)$
are hard constraints corresponding to the modulation and code constraints, respectively. 
Further factorization of these terms are available for common modulations and codes in the literature~\cite{Wymeersch2007}.
The factor graph of the joint distribution is shown in Fig.~\ref{fig:fg_orig}.
This model is representative of interference corrupted observations 
due to co-channel interference or a non-orthogonal multiple access scheme.
The developments presented with this model are applicable to other 
multi-signal or interference models.

The way in which the BP-MF framework is applied to the multi-signal scenario is 
determined by how the factor graph is partitioned into BP and MF subgraphs.
The model of Fig.~\ref{fig:fg_orig} shows the standard approach to  
factor graph construction and partitioning~\cite{Manchon2011,Badiu2012_coop}.  
The partitioning falls along the symbols variables as shown by the 
dashed line in Fig.~\ref{fig:fg_orig}.  
As a formal definition of this partitioning we have the following sets:
\begin{align}
	&\mathcal{A}_{\rm BP} = \{ f_{C_i} | i\in[1:N] \} \\
	&\mathcal{A}_{\rm MF} = \{ f_{Y_k} | k \in [0:K-1] \} \cup \{ f_{h_n} | n \in [1:N] \}  
	\label{eq:partition}
\end{align}
with the associated sets of variables given by
\begin{align}
	& \mathcal{I}_{BP} = \{ \vb{x}_1,\ldots,\vb{x}_N \} \cup \{ \vb{c}_1, \ldots, \vb{c}_N \} \cup \{ \vb{b}_1, \ldots, \vb{b}_N \} \\
	& \mathcal{I}_{MF} = \{ \vb{x}_1,\ldots,\vb{x}_N \} \cup \{ \vb{h} \}.
\end{align}

\section{Factor Graph Construction for BP-MF} \label{sec:construct}
In this section, we develop a new factor graph construction for application of
the BP-MF framework to multi-signal detection.  We begin by showing
the limitations of the standard approach.
A summary of notation is provided in Table~\ref{tab:notation} 
for the parameters of the messages used in the following sections.
The parameters of the messages have a subscript which identifies the associated 
variable and an arrow which identifies the associated message according to the
direction it is passed in Figs.~\ref{fig:simple_fg} and \ref{fig:simple_jnt_fg}.

\subsection{Standard BP-MF Application}
\begin{figure}
	\centering
		\includegraphics[width=0.3\columnwidth]{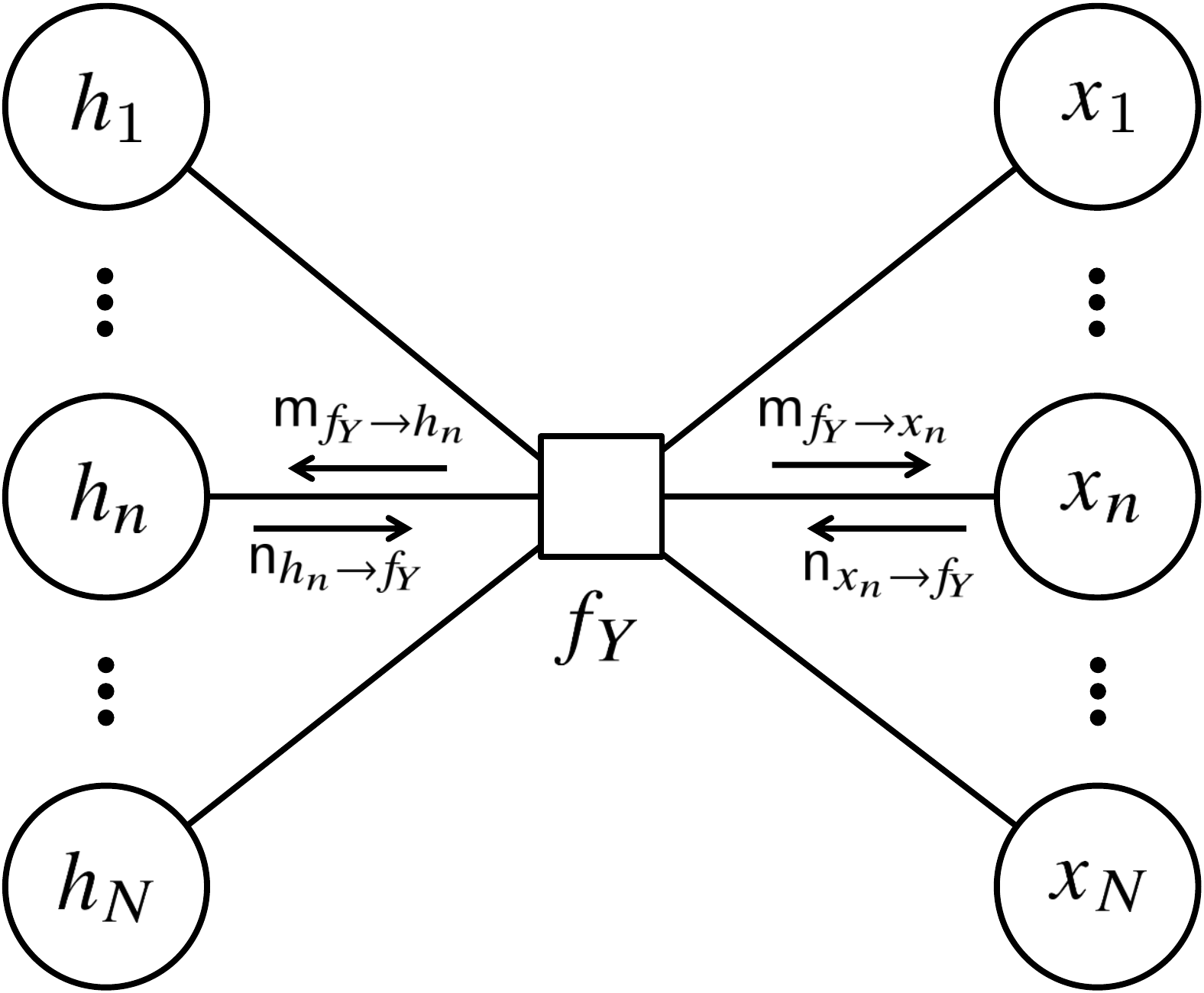}
		\caption{Factor graph of a \emph{typical} interference-corrupted observation model based on the standard BP-MF construction.}
		\label{fig:simple_fg}
\end{figure}

Based on the standard approach to applying BP-MF to multi-signal models,
a \emph{typical} observation factor is shown in Fig.~\ref{fig:simple_fg}
as a point of reference for the following work. 
The reference to subscript $k$ has been removed in order to simply the
notation.
The factor function for $f_Y$ in Fig.~\ref{fig:simple_fg} is the 
likelihood function which is given by
\begin{align}
	&p(y | \vb{x}, \vb{h}) = \frac{\gamma}{\pi} \exp \left( -\gamma \left| y - \sum_{n=1}^N h_n x_n \right|^2 \right) \notag \\
	&\ \propto \exp \left( -\gamma \left( - 2 \Re \left\{ y \sum_{n=1}^N h_n^* x_n^* \right\} + \sum_{n_1=1}^N \sum_{n_2=1}^N h_{n_1} h_{n_2}^* x_{n_1} x_{n_2}^* \right)  \right).
	\label{eq:simple_likelihood}
\end{align}

\begin{table}[htb]
	\centering
	\renewcommand{\arraystretch}{1.2}
  \caption{Summary of notation}
  \begin{tabular}{@{}ccccc@{}}
		\toprule
			Messages & Messages & Mean & Covariance & Correlation \\
			Fig.~\ref{fig:simple_fg} & Fig.~\ref{fig:simple_jnt_fg} \\
			\midrule
			$\mfg_{f_Y \rightarrow h_n}$ & $\mfg_{f_Q \rightarrow h_n}$ & $\paraml{\mu}_{h_n}$ & $\paraml{\sigma}^2_{h_n}$ \\	
			$\nfg_{h_n \rightarrow f_Y}$ & $\nfg_{h_n \rightarrow f_Q}$ & $\paramr{\mu}_{h_n}$ & $\paramr{\sigma}^2_{h_n}$ & $\paramr{\rho}_{h_{n}}$ \vspace{0.4em} \\	
			 & $\mfg_{f_Q \rightarrow \vb{q}}$ & $\paramR{\mu}_{\vb{q}}$ & $\paramR{\Sigma}_{\vb{q}}$ & $\paramR{\vb{R}}_{\vb{q}}$ \\	
			 & $\nfg_{\vb{q} \rightarrow f_Q}$ & $\paramL{\mu}_{\vb{q}}$ & $\paramL{\Sigma}_{\vb{q}}$ & $\paramL{\vb{R}}_{\vb{q}}$ \vspace{0.4em} \\	
			 & $\mfg_{f_Y \rightarrow \vb{q}}$ & $\paraml{\mu}_{\vb{q}}$ & $\paraml{\Sigma}_{\vb{q}}$ & $\paraml{\vb{R}}_{\vb{q}}$ \\	
			 & $\nfg_{\vb{q} \rightarrow f_Y}$ & $\paramr{\mu}_{\vb{q}}$ & $\paramr{\Sigma}_{\vb{q}}$ & $\paramr{\vb{R}}_{\vb{q}}$ \vspace{0.4em} \\
			 & $\mfg_{f_Y \rightarrow \vb{s}}$ & $\paramr{\mu}_{\vb{s}}$ & $\paramr{\Sigma}_{\vb{s}}$ & $\paramr{\vb{R}}_{\vb{s}}$ \\	
			 & $\nfg_{\vb{s} \rightarrow f_Y}$ & $\paraml{\mu}_{\vb{s}}$ & $\paraml{\Sigma}_{\vb{s}}$ & $\paraml{\vb{R}}_{\vb{s}}$ \vspace{0.4em} \\
			 & $\mfg_{f_S \rightarrow \vb{s}}$ & $\paramL{\mu}_{\vb{s}}$ & $\paramL{\Sigma}_{\vb{s}}$ & $\paramL{\vb{R}}_{\vb{s}}$ \\	
			 & $\nfg_{\vb{s} \rightarrow f_S}$ & $\paramR{\mu}_{\vb{s}}$ & $\paramR{\Sigma}_{\vb{s}}$ & $\paramR{\vb{R}}_{\vb{s}}$ \vspace{0.4em} \\
			$\mfg_{f_Y \rightarrow x_n}$ & $\mfg_{f_S \rightarrow x_n}$ & $\paramr{\mu}_{x_n}$ & $\paramr{\sigma}^2_{x_n}$ \\	
			$\nfg_{x_n \rightarrow f_Y}$ & $\nfg_{x_n \rightarrow f_S}$ & $\paraml{\mu}_{x_n}$ & $\paraml{\sigma}^2_{x_n}$ & $\paraml{\rho}_{x_{n}}$ \\	
		\bottomrule
  \end{tabular}
  \label{tab:notation}
\end{table}

According to the MF approximation, the messages from the 
observation factor to the channel coefficients are given by
\begin{equation}
	\mfg_{f_Y \rightarrow h_n}(h_n) = \exp \left( \idotsint \sum_{i=1}^{N} \sum_{x_i} \ln ( p(y | \vb{x},\vb{h}) ) \nfg_{x_i \rightarrow f_Y}(x_i) \prod_{j \neq n} \nfg_{h_{j} \rightarrow f_Y}(h_{j}) dh_{j} \right).
	\label{eq:simple_mfyh}
\end{equation}
The MF rule is less complex than BP because the expectation is taken on the argument of the exponential function due to $p(y|\vb{x},\vb{h})$ being in the exponential family.
After performing the expectations the message is proportional to a complex Gaussian distribution as given by
\begin{align}
	\mfg_{f_Y \rightarrow h_n}(h_n) &\propto \exp \left( -\gamma \left( \paraml{\rho}_{x_{n}} |h_n|^2 - 2 \Re \left\{ h_n^* \paraml{\mu}_{x_n}^* \left(y - \sum_{n^\prime \neq n} \paramr{\mu}_{h_{n^\prime}} \paraml{\mu}_{x_{n^\prime}} \right) \right\} \right) \right) \notag \\
		&\propto \mathcal{CN} \left( h_n ; \paraml{\mu}_{h_n} , \paraml{\sigma}^2_{h_n} \right),
	\label{eq:simple_mfyh_final}
\end{align}
where the mean and variance are $\paraml{\mu}_{h_n} = \paraml{\rho}_{x_{n}}^{-1} \paraml{\mu}_{x_n}^* \left(y - \sum_{n^\prime \neq n} \paramr{\mu}_{h_{n^\prime}} \paraml{\mu}_{x_{n^\prime}} \right)$ and $\paraml{\sigma}^2_{h_n} = (\gamma \paraml{\rho}_{x_{n}})^{-1}$, respectively.
Similarly, the messages from the observation factor to the symbols are given by
\begin{align}
		\mfg_{f_Y \rightarrow x_n}(x_n) &\propto \exp \left( -\gamma \left( \paramr{\rho}_{h_{n}} |x_n|^2 - 2\Re \left\{ x_n^* \paramr{\mu}_{h_n}^* \left(y - \sum_{n^\prime \neq n} \paramr{\mu}_{h_{n^\prime}} \paraml{\mu}_{x_{n^\prime}} \right) \right\} \right) \right) \notag \\
		&\propto \mathcal{CN} \left( x_n ; \paramr{\mu}_{x_n} , \paramr{\sigma}^2_{x_n} \right),
	\label{eq:simple_mfyx}
\end{align}
where the mean and variance are $\paramr{\mu}_{x_n} = \paramr{\rho}_{h_{n}}^{-1} \paramr{\mu}_{h_n}^* \left(y - \sum_{n^\prime \neq n} \paramr{\mu}_{h_{n^\prime}} \paraml{\mu}_{x_{n^\prime}} \right)$ and $\paramr{\sigma}^2_{x_n} = \gamma \paramr{\rho}_{h_{n}}$, respectively.
We conclude from \eqref{eq:simple_mfyh_final} and \eqref{eq:simple_mfyx} that the
MF approximation leads to an interference cancellation structure 
when computing the outgoing messages.  The uncertainty in the variables 
(i.e., the variance) is not considered in the interference cancellation part; 
only the mean is used for terms $n^\prime \neq n$ in \eqref{eq:simple_mfyh} 
and \eqref{eq:simple_mfyx}.
Interference cancellation, which only accounts for the mean,
are known to be inferior to approaches which use both the mean and variance 
of the incoming messages~\cite{Ping2006,Jiang2009}.
When estimation of the noise variance (or precision) 
is included in the MF framework, it naturally accounts for errors 
in the interference cancellation and, therefore, indirectly the uncertainty 
is accounted for.
However, a single variance parameter does not effectively capture the variance of 
each individual symbol when performing cancellation.
In simulations, we found that due to the interference cancellation structure, 
the MF approximation was a particularly poor
choice for equalization in the presence of interference.
This has motivated us to explore and propose an alternative model for 
multi-signal detection. 

\subsection{Joint Auxiliary Variables}
\begin{figure*}
	\centering
		\includegraphics[width=0.8\textwidth]{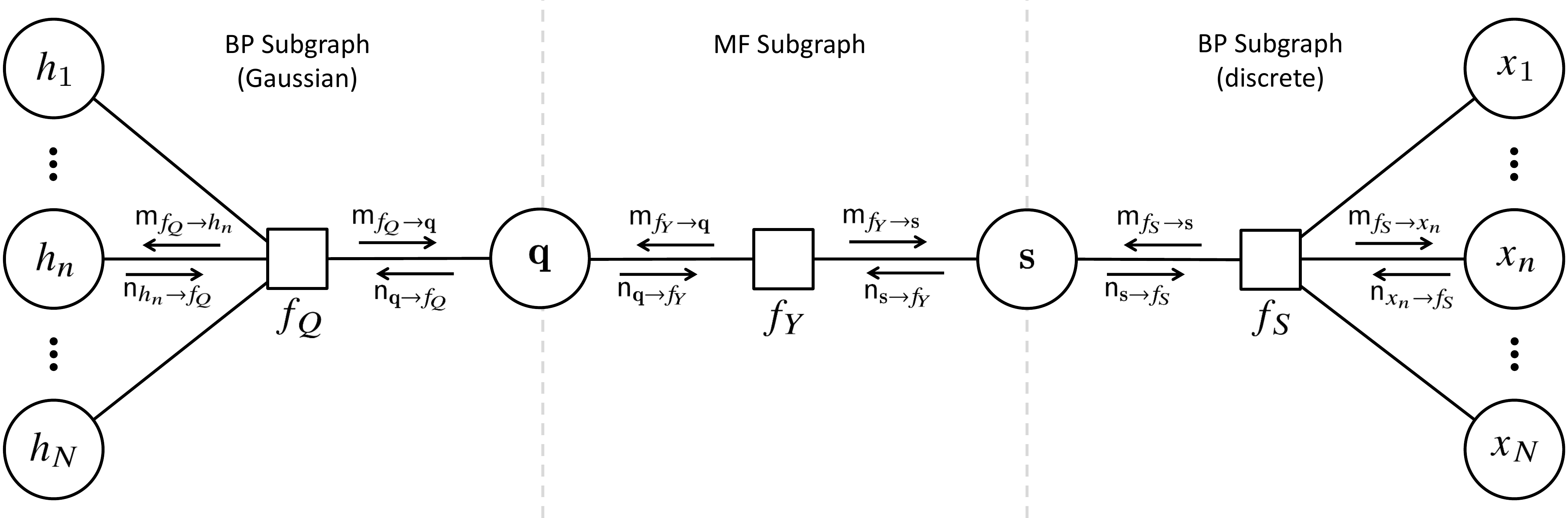}
		\caption{Factor graph of a \emph{typical} interference corrupted observation model based on the proposed BP-MF construction with auxiliary variables. In the graph,  
		$f_Q$ labels the factor $p(\vb{q}|h_1,\ldots,h_N)$ and
		$f_S$ labels the factor $p(\vb{s}|x_1,\ldots,x_N)$.}
		\label{fig:simple_jnt_fg}
\end{figure*}

The proposed factor graph model is based on separation of the equalization
and channel estimation functions from the observations.  The goal is to 
apply more effective BP for these functions while still maintaining the 
advantages of the MF approximation in regards to the non-linear observation model.
To separate the equalization and channel estimation functions, we introduce
two auxiliary variables:
a joint channel coefficient variable and a joint symbol variable as defined by
\begin{equation}
	\vb{s} \triangleq \vb{x} = [x_1,\ldots,x_N]^\Tra \ \ \text{and} \ \ \vb{q} \triangleq \vb{h} = [h_1,\ldots,h_N]^\Tra,
	\label{eq:defn_auxiliary}
\end{equation} 
respectively.
This enables us to factor the joint distribution into distinct observation, equalization, and channel estimation factors as given by
\begin{equation}
	p(y,\vb{x},\vb{h},\vb{s},\vb{q}) = \underbrace{p(y | \vb{s},\vb{q})}_{f_Y} \underbrace{p(\vb{s}|\vb{x})}_{f_S} \underbrace{p(\vb{q} | \vb{h})}_{f_Q} \prod_{i=1}^N p(x_i) \prod_{i=1}^N p(h_i).
\end{equation}
The factor graph model is shown in Fig.~\ref{fig:simple_jnt_fg}.

With the introduction of joint auxiliary variables, the MF subgraph
becomes a link between estimation of the channel coefficients
and equalization of the symbols.  
Summary statistics (i.e., the first and second order moments)
of the channel coefficient based on incoming message $\nfg_{\vb{q} \rightarrow f_Y}$
are used for detection of the data.  Similarly, first and second order moments of 
the symbols based on the incoming message $\nfg_{\vb{s} \rightarrow f_Y}$ are used for channel estimation.  
The key difference is that joint 
distributions for all symbols (or channel coefficients) flow away from 
the observation nodes.  Thus, the interference cancellation structure has 
been eliminated.  As we will show, this allows us to perform equalization
or channel estimation using belief propagation which is more successful. 
These points are demonstrated by the following message derivations.

The likelihood function is expressed in terms of the vector notation 
defined in \eqref{eq:defn_auxiliary} and is given by
\begin{align}
	p(y | \vb{s},\vb{q}) \propto \exp \left( - \gamma \left| y - \vb{q}^\Tra \vb{s} \right|^2 \right).
\end{align}
The message from the observation to the auxiliary symbol variable is
derived as follows:
\begin{align}
	&\mfg_{f_Y \rightarrow \vb{s}}(\vb{s}) \notag \\
		&\ \propto \exp \left\{ - \gamma \left( -y \vb{s}^\Hem \paramr{\mu}_{\vb{q}}^* - y^* \paramr{\mu}_{\vb{q}}^\Tra \vb{s} + \vb{s}^\Hem \paramr{\vb{R_q}} \vb{s} \right) \right\} \label{eq:simple_jnt_mys_noinv} \\
		&\ \propto \mathcal{CN} \left( \vb{s} ; \paramr{\mu}_{\vb{s}} , \paramr{\Sigma}_{\vb{s}} \right),
		\label{eq:simple_jnt_mys}
\end{align}
where $\paramr{\mu}_{\vb{s}} = \paramr{\vb{R_q}}^{-1} y \paramr{\mu}_{\vb{q}}^*$ and $\paramr{\Sigma}_{\vb{s}} = \left( \gamma\paramr{\vb{R_q}} \right)^{-1}$.
The message from the observation to the auxiliary channel variable
is derived similarly as given by
\begin{align}
	&\mfg_{f_Y \rightarrow \vb{q}}(\vb{q}) \notag \\
		&\ \propto \exp \left\{ - \gamma \left( -y \vb{q}^\Hem \paraml{\mu}_{\vb{s}}^* - y^* \paraml{\mu}_{\vb{s}}^\Tra \vb{q} + \vb{q}^\Hem \paramr{\vb{R_s}} \vb{q} \right) \right\} \notag \\
		&\ \propto \exp \left\{ - \gamma \left( \vb{q} - \paraml{\vb{R_s}}^{-1} y \paraml{\mu}_{\vb{s}}^* \right)^\Hem \paraml{\vb{R_s}} \left( \vb{q} - \paraml{\vb{R_s}}^{-1} y \paraml{\mu}_{\vb{s}}^* \right) \right\} \notag \\
		&\ \propto \mathcal{CN} \left( \vb{q} ; \paraml{\mu}_{\vb{q}} , \paraml{\Sigma}_{\vb{q}} \right),
\end{align}
where $\paraml{\mu}_{\vb{q}} = \paraml{\vb{R_s}}^{-1} y \paraml{\mu}_{\vb{s}}^*$ and $\paraml{\Sigma}_{\vb{q}} = \left( \gamma \paraml{\vb{R_s}} \right)^{-1} $.
Descriptions for the parameters can be found in Table~\ref{tab:notation}.

\subsection{BP-Based Channel Estimation} \label{sec:bp_channel}
In contrast to the interference cancellation structure of the MF approximation,
Gaussian BP follows the structure of LMMSE filtering where the
incoming messages ($\nfg_{h_n \rightarrow f_Q}(h_n)$) are treated as
prior distributions.  
We derive the messages associated with the channel estimation 
BP subgraph in Appendix~\ref{sec:channel_gbp}.
Rather than rely only on ``soft'' symbol estimates, 
the proposed algorithm makes use of 
first and second order moments of the symbols as a result of the MF approximation.

\subsection{BP-Based Equalization}
Because the joint auxiliary symbol variable has a discrete domain, the 
message $\mfg_{f_Y \rightarrow \vb{s}}(\vb{s})$ is a discrete distribution
which is computed by evaluating \eqref{eq:simple_jnt_mys_noinv} for all vectors $\vb{s}$. 
The sum-product rule is applied to messages from the factor $f_S$ to the neighboring
variables.  In this context, discrete BP (or the sum-product algorithm) is 
a (local) joint MAP detector where incoming messages from the symbols are treated 
as prior distributions.  We derive the messages associated with the
equalization BP subgraph in Appendix~\ref{sec:equal_spa}.

\section{Combined BP-MF-EP Receiver Architecture} \label{sec:approx}
In the previous section, we applied the BP-MF framework to detection of 
multiple signals.  The advancement in our work is a factor graph model
that maintains the benefits of the MF approximation and enables BP to be
used for equalization and channel estimation. 
The complexity of BP for discrete variables is a function of the variable's
domain size and number of signal components in the observation.
For example, if an observation is comprised of $N$ signal components
with domain $\mathcal{X}$, the complexity of BP is $\mathcal{O}(|\mathcal{X}|^N)$.
In other words, it is exponential in the number of components and number of bits per 
symbol (i.e., bits per signal component).
While this is not prohibitive for detection of several signals with 
low-order modulation 
(for example, 2 signals with QPSK modulation has a complexity order of 16), 
it is a barrier when more signal components are present or have high-order modulations
(for example, 4 signals with 16 QAM modulation has a complexity order of 65\,536)

We develop a reduced complexity receiver for the proposed model of 
Section~\ref{sec:construct} by making the assumption that the symbols
and the joint auxiliary symbol variable are continuous random variables
and that the messages returned from demodulation and decoding are
Gaussian distributed.  With these assumptions, Gaussian BP is applied 
to the equalization factor $f_S$ and associated variables.
The Gaussian BP algorithm for equalization is identical to Gaussian BP
for channel estimation.  
As with channel estimation, the Gaussian BP equalizer has the form of LMMSE 
filtering with prior information and makes use of first and second order moments
of the channel coefficients according to the MF messages. 
Gaussian BP has been applied to equalization in iterative receivers before
and has been shown to be equivalent to LMMSE filtering where feedback from
the decoder is treated as prior information~\cite{Guo2008}.

The question that remains is how to compute Gaussian distributed messages from 
the discrete messages returned from demodulation and decoding.  
In similar applications, a Gaussian distribution whose mean and variance 
match the mean and variance of the extrinsic distribution has been proposed~\cite{Rossi2008}. 
In some cases it was found to be more effective to match the mean 
and variance to the posterior discrete distribution~\cite{Witzke2002}.
Senst and Ascheid provided a theoretically justified approach to computing the
Gaussian messages based on EP~\cite{Senst2011}. 
Their work also provides insight into why posterior distributions are more effective
than extrinsic distributions.  

\subsection{Gaussian Approximation}
The joint auxiliary symbol variable is approximated as having a continuous 
domain. Similarly, the symbols $x_1,\ldots, x_N$ are approximated as having
continuous domains for factor $f_S$ (i.e., $p(\vb{s}|x_1,\ldots,x_N)$) and
maintain a discrete domain in the factors $f_{C_1},\ldots,f_{C_N}$.
The messages from the symbols to the equalization factors $\nfg_{x_n \rightarrow f_S}(x_n)$
are now continuous and are approximated as Gaussian distributions.
Gaussian BP is used to compute messages involving the equalization factor
in a similar manner to Section~\ref{sec:bp_channel} and Appendix~\ref{sec:channel_gbp}.  

\subsection{Expectation Propagation}
From Fig.~\ref{fig:high_lvl_fg}, messages $\mfg_{f_S \rightarrow x_n}(x_n)$ and $\nfg_{x_n \rightarrow f_S}(x_n)$
are Gaussian distributions while 
$\mfg_{f_{C_n} \rightarrow x_n}(x_n)$ and $\nfg_{x_n \rightarrow f_{C_n}}(x_n)$
are discrete distributions.  Computing $\nfg_{x_n \rightarrow f_{C_n}}(x_n)$
from $\mfg_{f_S \rightarrow x_n}(x_n)$ is straightforward: the Gaussian distribution 
is evaluated for each value in the domain of the symbol.  In order to approximate 
discrete distribution $\mfg_{f_{C_n} \rightarrow x_n}(x_n)$ with Gaussian 
distribution $\nfg_{x_n \rightarrow f_S}(x_n)$, we apply EP.  

EP is implemented by first computing the exact belief from the discrete distributions
as given by
\begin{equation}
	b(x_n) = \mfg_{f_{C_n} \rightarrow x_n}(x_n) \nfg_{x_n \rightarrow f_{C_n}}(x_n).
	\label{eq:xn_belief}
\end{equation}
Subsequently, the belief is approximated with a Gaussian distribution as given by
\begin{equation}
	\tilde{b}(x_n) = \mathcal{CN}(x_n; \mu_{x_n} , \sigma^2_{x_n}) ,
\end{equation}
where $\mu_{x_n}$ and $\sigma^2_{x_n}$ \emph{without arrows} denote parameters of the belief
computed from \eqref{eq:xn_belief}.
Finally, the parameters of the Gaussian distribution for $\nfg_{x_n \rightarrow f_S}(x_n)$ 
are computed by dividing $\tilde{b}(x_n)$ by $\mfg_{f_S \rightarrow x_n}(x_n)$.
Thus, the mean and variance of $\nfg_{x_n \rightarrow f_S}(x_n)$ are given by
\begin{equation}
	\paraml{\sigma}^2_{x_n} = \left( \frac{1}{\sigma^2_{x_n}} - \frac{1}{\paramr{\sigma}^2_{x_n}} \right)^{-1}
	\label{eq:ep_var}
\end{equation}
and
\begin{equation}
	\paraml{\mu}_{x_n} = \paraml{\sigma}^2_{x_n} \left( \frac{\mu_{x_n}}{\sigma^2_{x_n}} - \frac{\paramr{\mu}_{x_n}}{\paramr{\sigma}^2_{x_n}} \right),
\end{equation}
respectively.
It is possible that the variance computed in \eqref{eq:ep_var} is negative.  
In this case, we apply the solution used in~\cite{Senst2011} and approximate
the message by the Gaussian belief, i.e., we set
$\paraml{\mu}_{x_n} = \mu_{x_n}$ and $\paraml{\sigma}^2_{x_n} = \sigma^2_{x_n}$.

\subsection{Receiver Architecture}
\begin{figure*}
	\centering
		\includegraphics[width=0.75\textwidth]{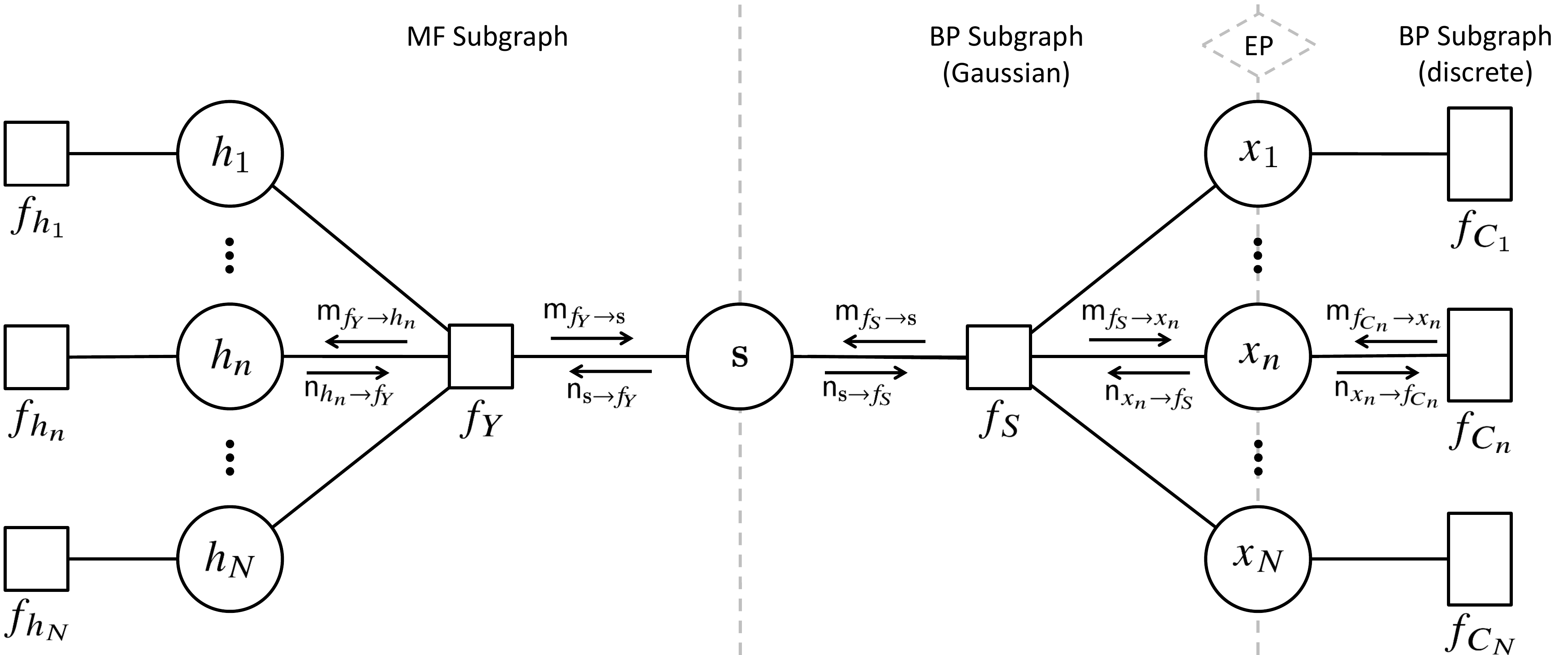}
		\caption{Factor graph of a \emph{typical} observation based on the proposed BP-MF-EP receiver architecture for low-complexity implementation.}
		\label{fig:high_lvl_fg}
\end{figure*}

In summary we propose a receiver architecture for multi-signal detection 
which combines Gaussian and discrete BP, the MF approximation, and EP.
The architecture is built upon the factor graph model of the joint distribution
including choice of auxiliary variables and partitioning of the graph into
subgraphs.  We divide the graph into three regions as follows:
\begin{itemize}
	\item MF subgraph including the observation and channel estimation factors
	and corresponding variables. 
	\item Gaussian BP subgraph including the equalization factors, and
	\item Discrete BP subgraph including the modulation and coding constraint
	factors.
\end{itemize}
Channel estimation can be separated into its own subgraph
applying Gaussian BP or GAMP.  Additionally, channel estimation can be 
accomplished using the EM algorithm as a special case of the MF approximation.
A diagram of the receiver architecture is shown in Fig.~\ref{fig:high_lvl_fg}
In the next section, we demonstrate this receiver architecture by applying it
to MIMO-OFDM.

\section{Application to MIMO-OFDM} \label{sec:der_mimo}
In this section, the proposed receiver architecture is applied to reception of
MIMO-OFDM signals.  
This model can be used to accomplish multi-signal detection in general
(e.g., single antenna multiuser detection or multiuser MIMO schemes).

\subsection{System Model}
Consider a MIMO-OFDM transmission scheme 
which utilizes $N$ transmit antennas, $M$ receive antennas, 
and $K$ subcarriers.
The information bits, coded bits, and symbols for the $i$th stream
transmitted on the $i$th antenna (similarly, the $i$th user)
are denoted by $\vb{b}_i$, $\vb{c}_i$, and $\vb{x}_i$, respectively.
The symbols transmitted on the $k$th subcarrier across all antennas 
are collected into vector $\vb{x}(k) = [x_1(k), \ldots, x_N(k)]^\Tra$.

The multipath channel between each pair of transmitter and receiver 
antennas is modeled with a conventional tapped delay line with $L$ taps spaced
at the OFDM symbol sample rate.  The channel coefficients associated with
the $n$th transmitter antenna and the $m$th receiver antenna are denoted by the $L \times 1$ vector
$\vb{h}_{mn} = [h_{mn}(0), h_{mn}(1), \ldots, h_{mn}(L-1)]^\Tra$.
The collection of the channel coefficients for all pairs of transmitter and 
receiver antennas is given by
\[
	\vb{h} = \left[ \left[\vb{h}_{11}^\Tra, \ldots, \vb{h}_{M1}^\Tra \right], \ldots, 
			\left[\vb{h}_{1N}^\Tra, \ldots, \vb{h}_{MN}^\Tra \right] \right]^\Tra.
	\]
The channel coefficients $h_{mn}(l)$ for all $m=1,\ldots,M$, $n=1,\ldots,N$, 
and $l=0,\ldots,L-1$ are assumed to be independent.

The frequency domain channel coefficients for the $K$ subcarriers are 
obtained through the Fourier transform.  We define a $K \times L$ DFT 
matrix $\vb{D}$ where the $k,l$th element is given by
$d_{kl} = e^{-j2\pi k l /K}$.
The frequency domain channel coefficients are defined as given by
\begin{equation}
	\tilde{h}_{mn}(k) = \sum_{l=0}^{L-1} h_{mn}(l) d_{kl} 
	\label{eq:channel_individual}
\end{equation} 
for all $m=1,\ldots,M$, $n=1,\ldots,N$, and $k=0,\ldots,K-1$.
The $M \times N$ MIMO channel matrix for the $k$th subcarrier is given by
\begin{equation}
	\tilde{\vb{H}}(k) = 
		\begin{bmatrix} 
			\tilde{h}_{11}(k) & \tilde{h}_{12}(k) & \cdots & \tilde{h}_{1N}(k) \\
			\tilde{h}_{21}(k) & \tilde{h}_{22}(k) &  & \vdots \\
			\vdots & & \ddots \\
			\tilde{h}_{M1}(k) & \cdots &  & \tilde{h}_{MN}(k)
		\end{bmatrix}.
\end{equation}

We define both scalar and vector forms for the received signal as follows.
The $M\times 1$ received signal vector for the $k$th subcarrier is given by
\begin{equation}
	\vb{y}(k) = \tilde{\vb{H}}(k) \vb{x}(k) + \vb{w}(k) \quad \forall k=0,1,\ldots, K-1 ,
	\label{eq:signal_vector}
\end{equation}
where and $\vb{w}(k)$ is a $M\times 1$ vector of iid 
circularly symmetric complex Gaussian random variables representing noise.
The per antenna noise precision is denoted by $\gamma$ (i.e., the noise
variance is given by $1/\gamma$).
The scalar form of the received signal is given by
\begin{equation} 
	y_m(k) = \sum_{n=1}^N x_n(k) \sum_{l=0}^{L-1} h_{mn}(l) d_{kl} + w_m(k),
	\label{eq:signal_scalar}
\end{equation}
where $y_m(k)$ and $w_m(k)$ are the $m$th elements of $\vb{y}(k)$ and 
$\vb{w}(k)$, respectively.
The concatenation of the received signal vectors and channel matrices for all subcarriers
are denoted as $\vb{y} = [\vb{y}(0)^\Tra, \ldots, \vb{y}(K-1)^\Tra]^\Tra$
and
$\tilde{\vb{H}} = [\tilde{\vb{H}}(0)^\Tra, \ldots, \tilde{\vb{H}}(K-1)^\Tra]^\Tra$, respectively.

\subsection{Factor Graph}
\begin{figure*}[t]
	\centering
		\includegraphics[width=0.75\columnwidth]{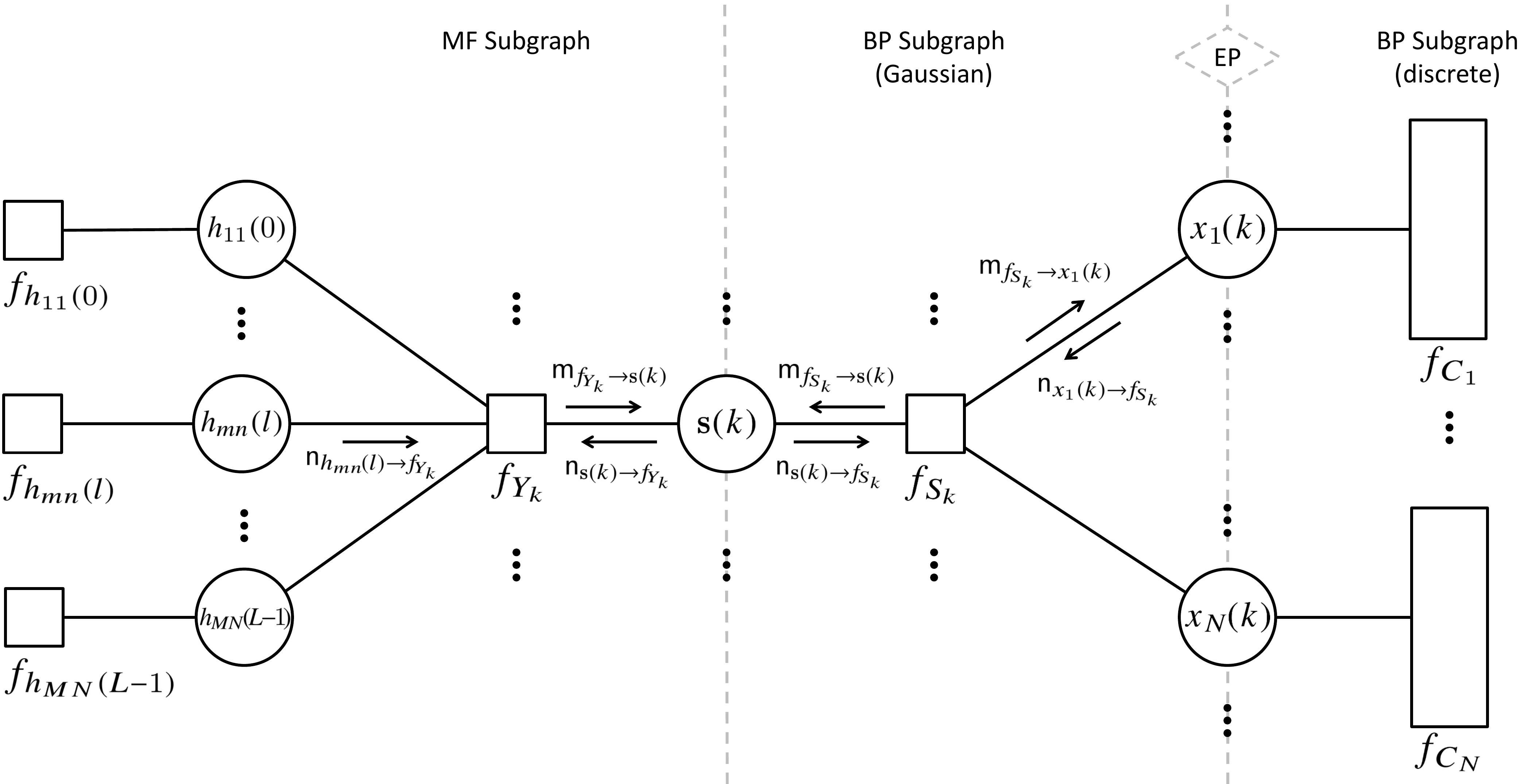}
		\caption{Factor graph of \eqref{eq:joint_sep} with auxiliary variables 
			$\vb{s}(k)$ enabling separation between channel estimation and detection.}
		\label{fig:fg_sep}
\end{figure*}

We introduce an auxiliary variable $\vb{s}(k)$ for each observation $\vb{y}(k)$ 
which represents the joint symbol vector for that observation 
(i.e., $x_1(k),\ldots,x_N(k)$).
The collection of all auxiliary variables is denoted as $\vb{s} = [\vb{s}(0)^\Tra,\ldots,\vb{s}(K-1)^\Tra]^\Tra$.
The joint distribution is factored as follows:
\begin{align}
	&p(\vb{y}, \vb{s}, \vb{x}_1, \ldots, \vb{x}_N,\vb{c}_1, \ldots, \vb{c}_N,\vb{b}_1, \ldots, \vb{b}_N,\vb{h}) \notag \\
	&\ = \prod_{k=0}^{K-1} p(\vb{y}(k) | \vb{s}(k) , \vb{h} ) p(\vb{s}(k) | x_1(k), \ldots, x_N(k)) \prod_{i=1}^{N} p(\vb{x}_i,\vb{c}_i,\vb{b}_i) \prod_{m=1}^M\prod_{n=1}^N\prod_{l=0}^{L-1}p(h_{mn}(l)).
	\label{eq:joint_sep}
\end{align}
The distributions $p(\vb{x}_i, \vb{c}_i, \vb{b}_i)$ may be further factored 
based on the modulation and code constraints which has been considered 
extensively in past work~\cite{Wymeersch2007}.  The factor graph of the joint
distribution is shown in Fig.~\ref{fig:fg_sep}. 

Two receiver architectures are constructed by defining the partitioning of the 
factor into subgraphs according to the message passing algorithm to be applied.
First, the \emph{exact} implementation applies BP-MF as constructed in 
Section~\ref{sec:construct}.
The exact implementation is defined by the following subgraphs:
\begin{align}
	&\mathcal{A}_{\rm BP} = \{ f_{S_k} | k \in [0:K-1] \} \cup \{ f_{C_i} | i\in[1:N] \} \label{eq:subgraphs_exact_start} \\
	&\mathcal{A}_{\rm MF} = \{ f_{Y_k} | k \in [0:K-1] \} \cup \{ f_{h_{mn}(l)} | m \in [1:M], n \in [1:N], l \in [0:L-1]  \} \\
	& \mathcal{I}_{BP} = \{ \vb{s}(k) | k \in [0:K-1] \} \cup \{ \vb{x}_1,\ldots,\vb{x}_N \} \cup \{ \vb{c}_1, \ldots, \vb{c}_N \} \cup \{ \vb{b}_1, \ldots, \vb{b}_N \} \\
	& \mathcal{I}_{MF} = \{ \vb{s}(k) | k \in [0:K-1] \} \cup \{ \vb{h} \}.
	\label{eq:subgraphs_exact_end}
\end{align}
Second, the approximate implementation applies BP-MF-EP as constructed in 
Section~\ref{sec:approx}. 
The approximate implementation is defined by the following subgraphs:
\begin{align}
	&\mathcal{A}_{\rm dBP} = \{ f_{C_i} | i\in[1:N] \} \label{eq:subgraphs_approx_start} \\
	&\mathcal{A}_{\rm GBP} = \{ f_{S_k} | k \in [0:K-1] \} \\
	&\mathcal{A}_{\rm MF} = \{ f_{Y_k} | k \in [0:K-1] \} \cup \{ f_{h_{mn}(l)} | m \in [1:M], n \in [1:N], l \in [0:L-1]  \} \\
	& \mathcal{I}_{dBP} = \{ \vb{x}_1,\ldots,\vb{x}_N \} \cup \{ \vb{c}_1, \ldots, \vb{c}_N \} \cup \{ \vb{b}_1, \ldots, \vb{b}_N \} \\
	& \mathcal{I}_{GBP} = \{ \vb{s}(k) | k \in [0:K-1] \} \cup \{ \vb{x}_1,\ldots,\vb{x}_N \} \\
	& \mathcal{I}_{MF} = \{ \vb{s}(k) | k \in [0:K-1] \} \cup \{ \vb{h} \},
	\label{eq:subgraphs_approx_end}
\end{align}
where $\mathcal{A}_{\rm dBP}$ denotes the discrete BP subgraph and $\mathcal{A}_{\rm GBP}$ denotes the Gaussian BP subgraph.
We handle channel estimation differently than previous work~\cite{Riegler2013,Badiu2012bpmf,Badiu2013,Yuan2016arxiv} by estimating the time-domain channel taps with the MF approximation.  Derivations for the messages within the MF subgraph are provided in Appendix~\ref{sec:mf_der}.

\section{Numerical Results} \label{sec:results}
In this section, we provide numerical results for the MIMO-OFDM receiver
using Monte Carlo simulation in order to validate the proposed 
probabilistic receiver architecture.
The simulation parameters are summarized in Table~\ref{tab:sim_param}.
Although internal loops are present within the factor graph, we do not
perform any sub-iterations within a full iteration of the receiver.
\begin{table}[htbp]
	\centering
	\renewcommand{\arraystretch}{1.2}
  \caption{Summary of the MIMO-OFDM simulation parameters}
  \begin{tabular}{@{}cc@{}}
		\toprule
			Parameter & Description \\
			\midrule
			Transmit antennas ($N$) & 4 \\
			Receive antennas ($M$) & 4 \\
			Subcarriers ($K$) & 300 \\
			OFDM Symbols/packet & 7 \\
			Coding & 1/2 PCCC \\
			Modulation & QPSK \\
			Reference Signals & 3GPP LTE (antenna ports 0--3) \\
			Channel taps ($L$) & 10 	\\		
		\bottomrule
  \end{tabular}
  \label{tab:sim_param}
\end{table}

We present simulation results for four receiver algorithms.
In all four cases, we use our proposed MF-based MIMO-OFDM channel estimation
developed in Section~\ref{sec:der_mimo} and Appendix~\ref{sec:mf_der}. 
The algorithms differ in how the BP-MF framework is applied to multi-signal detection.  
A description of each algorithm and the complexity of the equalization function is provided as follows:
\begin{enumerate}[(a)]
	\item \emph{BP-MF original}: the BP-MF implementation found in the prior art in which multi-signal detection is included in the MF subgraph (e.g., \cite{Manchon2011,Badiu2012_coop}).  The equalization function has a computational complexity $\mathcal{O}(N)$ as shown in \eqref{eq:simple_mfyx}.
	\item \emph{BP-MF exact}: our proposed BP-MF implementation in which discrete BP is applied to multi-signal detection (subgraphs defined in \eqref{eq:subgraphs_exact_start}--\eqref{eq:subgraphs_exact_end}).  The computational complexity of discrete BP for equalization of multiple signals is $\mathcal{O}(|\mathcal{X}|^N)$ where $|\mathcal{X}|$ is the modulation order.
	\item \emph{BP-MF-EP}: our proposed low-complexity BP-MF implementation with Gaussian BP and Gaussian approximations based on EP (subgraphs defined in \eqref{eq:subgraphs_approx_start}--\eqref{eq:subgraphs_approx_end}).  The computational complexity is $\mathcal{O}(N^3)$ due to the matrix inversion required for equalization.
	\item \emph{BP-MF approximate (ext)}: for comparison purposes a BP-MF implementation with Gaussian BP equalization and Gaussian approximations based on extrinsic distributions (subgraphs defined in \eqref{eq:subgraphs_approx_start}--\eqref{eq:subgraphs_approx_end}).  The computational complexity is also $\mathcal{O}(N^3)$.
\end{enumerate}

The performance with known channel coefficients and joint MAP (JMAP) detection
is also simulated to provide a lower bound. 
The JMAP receiver with known channel coefficients applies discrete BP and is 
identical to the \emph{BP-MF exact} receiver in its equalization, demodulation, 
and decoding functions.

\begin{figure}
	\centering
	\begin{minipage}[t]{0.48\textwidth}
		\centering
		\includegraphics[width=1.0\linewidth]{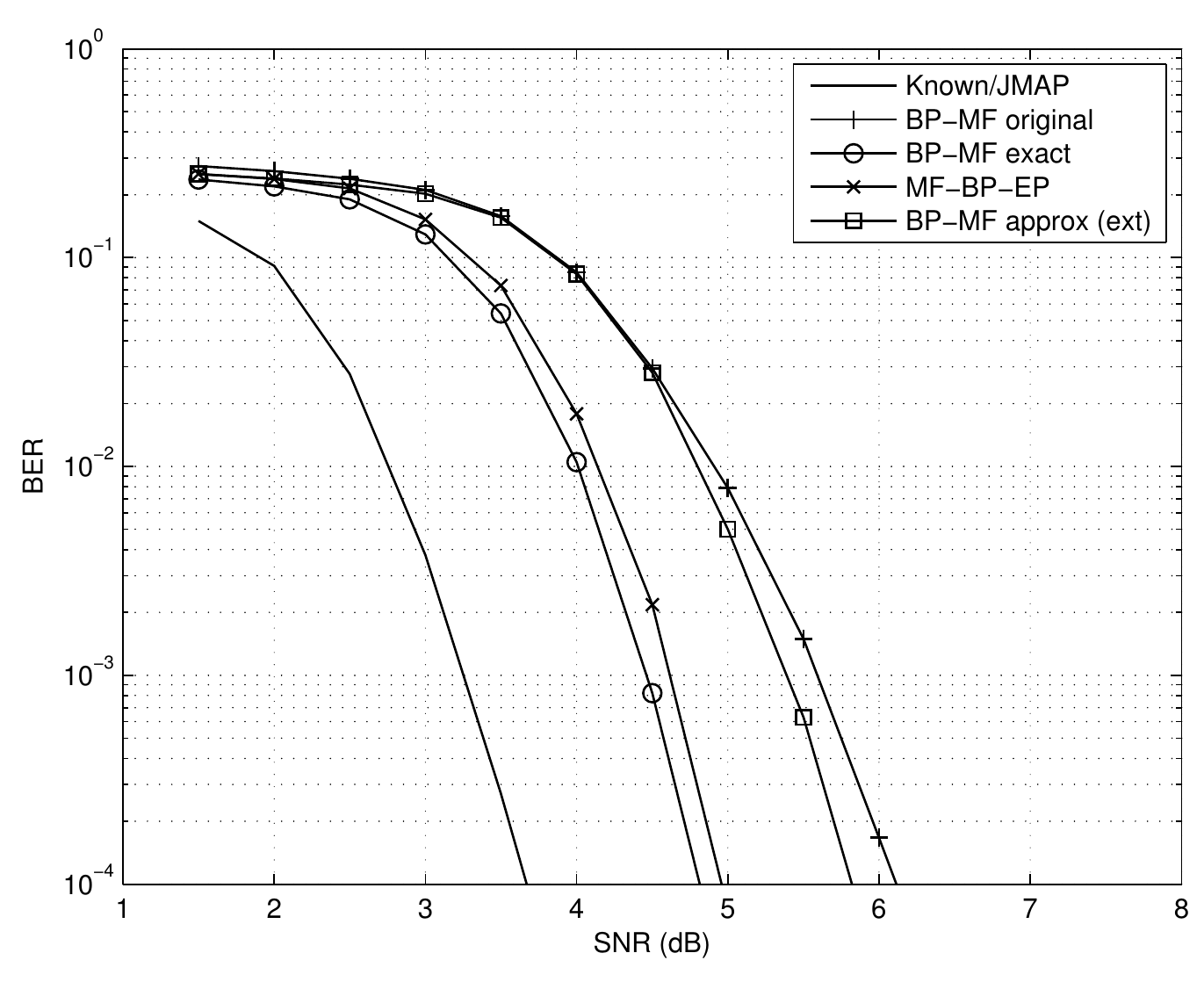}
		\caption{BER vs. SNR per antenna with 20 iterations.}
		\label{fig:ber_snr_mimo}	
	\end{minipage} \hfill%
	\begin{minipage}[t]{0.48\textwidth}
		\centering
		\includegraphics[width=1.0\linewidth]{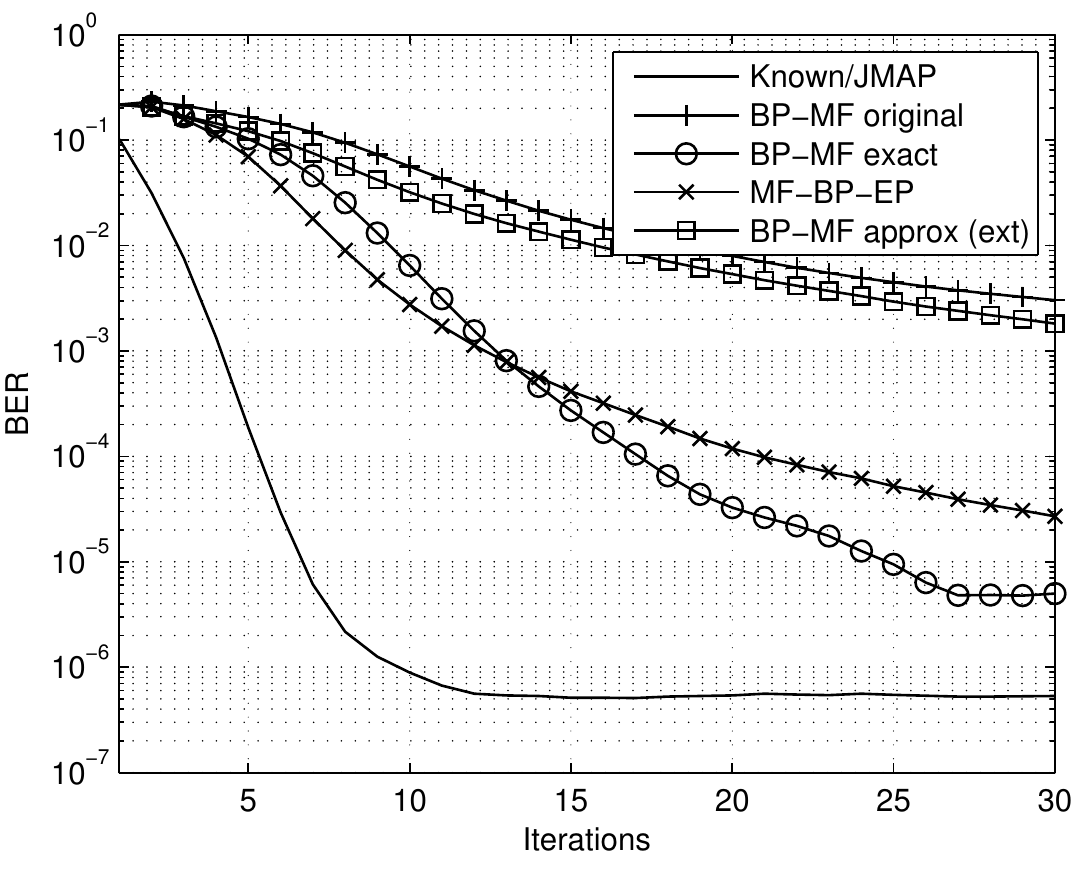}
		\caption{BER vs. iteration for $\text{SNR}=5$ dB per antenna.}
		\label{fig:ber_iter_mimo}
	\end{minipage}
\end{figure}

The bit error rate (BER) performance is shown with respect to SNR and with respect to the number of receiver iterations in Figs.~\ref{fig:ber_snr_mimo}
and \ref{fig:ber_iter_mimo}, respectively.  
The results support our claim that the interference cancellation structure 
of the MF approximation (used in the standard implementation) is not ideal 
for handling multi-signal detection.  This is because variance of a symbol's messages
(i.e., the degree of uncertainty about a particular symbol's value) is not accounted for
when applying the MF approximation to multi-signal detection.  
Applying BP to the MIMO equalization task (as done in the exact implementation of our proposed BP-MF construction) yields more than 1 dB improvement in performance and is about 1 dB away from the best achievable performance with perfect parameter knowledge.
However, our exact implementation of BP-MF comes at the cost of a computational complexity
which is exponential in the number of signals. 
The performance of the receivers which use Gaussian BP lies in between these two.
Specifically, Gaussian BP with extrinsic-based Gaussian approximations
does not significantly improve performance versus the \emph{BP-MF original}.
On the other hand, because EP is effective in computing Gaussian messages from
the discrete distributions passed to the Gaussian BP subgraph, the \emph{BP-MF-EP} receiver performs
very close to the \emph{BP-MF exact} receiver with discrete BP.  
Thus, with polynomial complexity order (like \emph{BP-MF original}), we are able to achieve nearly the performance of the exact implementation.
In fact, we observe in Fig.~\ref{fig:ber_iter_mimo}, that for iterations 1--13, \emph{BP-MF-EP} slightly out-performs the exact implementation.
This is likely due to the fact that \emph{BP-MF-EP} does not follow a strict use of extrinsic information which leads to faster convergence but an increased chance of ``hardening'' the distributions toward the wrong decisions (as seen in the cross-over of their BER curves in Fig.~\ref{fig:ber_iter_mimo}).

\begin{figure}
	\centering
	\begin{minipage}[t]{0.48\textwidth}
		\centering
		\includegraphics[width=1.0\linewidth]{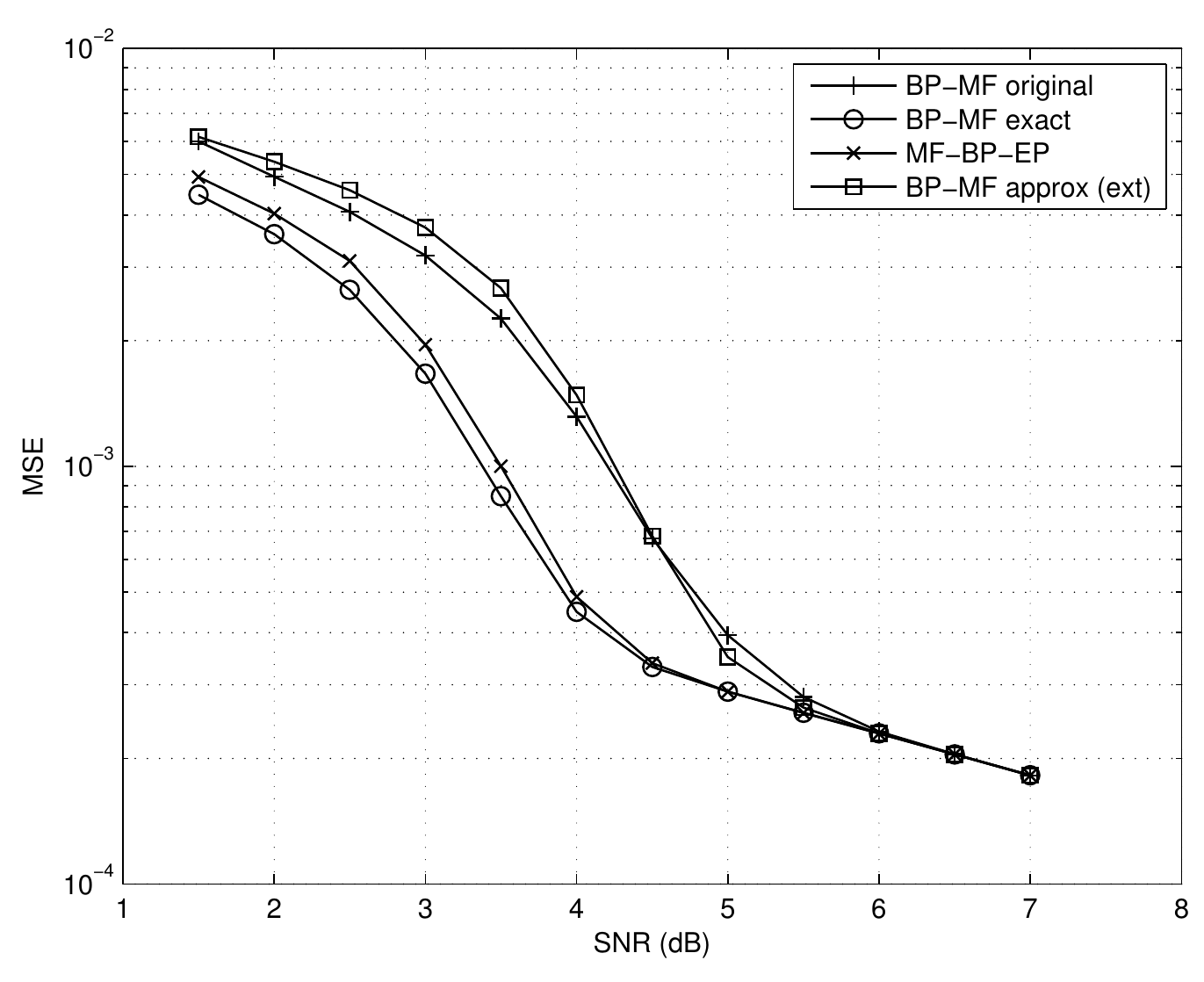}
		\caption{Channel estimation MSE vs. SNR per antenna with 20 iterations.}
		\label{fig:mse_snr_mimo}
	\end{minipage} \hfill%
	\begin{minipage}[t]{0.48\textwidth}
		\centering
		\includegraphics[width=1.0\linewidth]{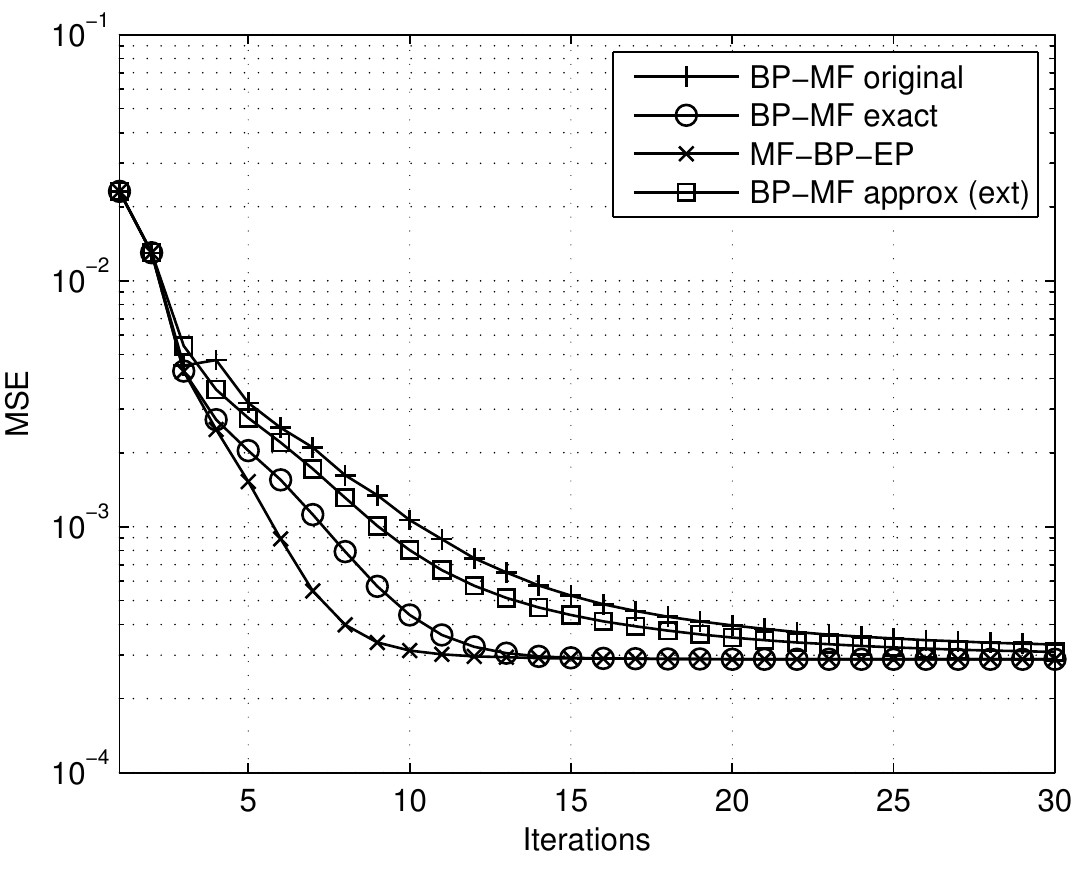}
		\caption{Channel estimation MSE vs. iteration for $\text{SNR}=5$ dB per antenna.}
		\label{fig:mse_iter_mimo}
	\end{minipage}
\end{figure}

The mean square error (MSE) channel estimation performance is shown in
Figs.~\ref{fig:mse_snr_mimo} and \ref{fig:mse_iter_mimo}. 
The only unexpected result is that, at lower SNR, the standard BP-MF implementation outperforms
Gaussian BP with extrinsic-based Gaussian approximations. 
This is further evidence that basing the Gaussian messages on the 
extrinsic distributions is a poor approximation.
As all receiver algorithms perform channel estimation in the same way, the difference in channel estimation performance is a consequence of the quality of the information about the data in each receiver.
The BP-MF-EP receiver again converges more quickly than the exact BP-MF receiver.  However, both receiver algorithms converge to the same estimation performance after about 12 iterations. 

%
%
%


\section{Conclusion} \label{sec:conclusion}
In this paper, we have developed a probabilistic receiver architecture 
for detection of multiple signals based on the BP-MF framework.  
By introducing 
auxiliary variables into the factor graph model, we maintain the benefits
of the MF approximation while avoiding an undesirable interference 
cancellation structure.
In scenarios in which the complexity of discrete BP for equalization is prohibitive,
we proposed Gaussian BP for multi-signal detection and a 
combined BP-MF-EP message passing algorithm.  
The proposed low-complexity algorithm is shown to perform nearly as well as 
the exact implementation for a MIMO-OFDM signal detection.
As a result of this work, we have developed a probabilistic receiver architecture with strong theoretical justification which can be applied 
to multi-signal detection and, in general, detection in the presence of 
interference.  
We have also developed a new MF-based time-domain channel estimation 
approach for MIMO-OFDM. While we have focused on MF-based channel estimation,
the factor graph construction also enables BP, GAMP, or EM to be applied to channel estimation.

\appendices

\section{Gaussian BP for Channel Estimation} \label{sec:channel_gbp}
The factor $f_Q$ is within a BP subgraph.  Thus, the message passed from the
joint auxiliary channel variable to this factor is an extrinsic message
as given by
\begin{equation}
	\nfg_{ \vb{q} \rightarrow f_Q}(\vb{q}) = \mfg_{f_Y \rightarrow \vb{q}}(\vb{q}).
\end{equation} 

Similarly, the messages from the channel coefficient variables to $f_Q$ are 
extrinsic messages which carry the prior distributions for the channel 
coefficients.  The factor $f_Q$ enforces equality between the channel 
coefficients and the corresponding terms within the joint auxiliary variable.
The factor function is given by the following hard constraint:
\begin{equation}
	p(\vb{q} | \vb{h}) = \prod_{i=1}^N \IndFn(q_i = h_i).
\end{equation}
When the prior distributions are Gaussian (e.g., for Rayleigh fading channels),
Gaussian BP is used to perform the computations for this factor node.

The joint auxiliary variable $\vb{q}$ is a concatenation of the channel 
coefficients. Thus, the BP message from $f_Q$ to the joint auxiliary variable 
is given by a concatenation of the input messages from the channel coefficients.
That is, the message is given by 
\begin{equation}
	\mfg_{f_Q \rightarrow \vb{q}}(\vb{q}) = \mathcal{CN}\left( \vb{q} ; \paramR{\mu}_{\vb{q}} , \paramR{\Sigma}_{\vb{q}} \right),
\end{equation}
where 
\begin{equation}
	\paramR{\mu}_{\vb{q}} = [\paramr{\mu}_{h_1},\ldots,\paramr{\mu}_{h_N}]^\Tra
\end{equation}
and
\begin{equation}
	\paramR{\Sigma}_{\vb{q}} = \text{diag} \left( \paramr{\sigma}^2_{h_1},\ldots,\paramr{\sigma}^2_{h_N} \right).
\end{equation}

The message from $f_Q$ to the channel coefficients is computed according 
to the sum-product rule.  In order to work with vector notation, the incoming
messages from the channel coefficients are combined into a mean and covariance
matrix as given by
\begin{equation}
	\mu_{\sim h_n} = [\paramr{\mu}_{h_1},\ldots,\paramr{\mu}_{h_{n-1}},0,\paramr{\mu}_{h_{n+1}},\ldots,\paramr{\mu}_{h_N}]^\Tra
\end{equation}
and
\begin{equation}
	V_{\sim h_n} = \text{diag} \left( \frac{1}{\paramr{\sigma}^2_{h_1}},\ldots, \frac{1}{\paramr{\sigma}^2_{h_{n-1}}}, 0, \frac{1}{\paramr{\sigma}^2_{h_{n+1}}}, \ldots, \frac{1}{\paramr{\sigma}^2_{h_N}} \right),
\end{equation}
where $h_n$ is excluded due to the sum-product rule.
With these definitions, the sum-product computation is as follows:
\begin{align}
	&\mfg_{f_Q \rightarrow h_n}(h_n) = \idotsint p(\vb{q} | \vb{h}) \nfg_{\vb{q} \rightarrow f_Q}(\vb{q}) \prod_{i \neq n} \nfg_{h_i \rightarrow f_Q}(h_i) dh_i d\vb{q}. \notag \\
		&\ \propto \idotsint \exp \left\{ - \left( \vb{h} - \paramL{\mu}_{\vb{q}} \right)^\Hem \paramL{\Sigma}_{\vb{q}}^{-1} \left( \vb{h} - \paramL{\mu}_{\vb{q}} \right) \right\} \notag \\
			&\ \qquad \cdot \exp \left\{ - \left( \vb{h} - \mu_{\sim h_n} \right)^\Hem V_{\sim h_n} \left( \vb{h} - \mu_{\sim h_n} \right) \right\}  \prod_{i \neq n} dh_i \notag \\
			&\ \propto \mathcal{CN}\left( h_n; \paraml{\mu}_{h_n} , \paraml{\sigma}^2_{h_n} \right) ,
	\label{eq:bp_channel}
\end{align}
where
\begin{equation}
	\paraml{\sigma}^2_{h_n} = \left[ \left( \paramL{\Sigma}_{\vb{q}}^{-1} + V_{\sim h_n} \right)^{-1} \right]_{n,n}
\end{equation}
and 
\begin{equation}
	\paraml{\mu}_{h_n} = \left[ \left( \paramL{\Sigma}_{\vb{q}}^{-1} + V_{\sim h_n} \right)^{-1} \left( \paramL{\Sigma}_{\vb{q}}^{-1} \paramL{\mu}_{\vb{q}} + V_{\sim h_n} \mu_{\sim h_n} \right)  \right]_{n}.
\end{equation}

An efficient implementation is to compute the joint posterior, marginalize,
and remove the input distribution to obtain marginal extrinsic messages
for each channel coefficient.  The joint posterior is computed once for the messages 
to all channel coefficient variables and, therefore, a single matrix inversion is
required.


Finally, the message from the joint channel auxiliary variable to the
observation factor is the posterior distribution for $\vb{q}$ since
$f_Y$ is in the MF subgraph.  The posterior is given by
\begin{align}
	\nfg_{\vb{q} \rightarrow f_Y}(\vb{q}) &= \mfg_{f_Q \rightarrow \vb{q}}(\vb{q}) \mfg_{f_Y \rightarrow \vb{q}}(\vb{q}) \notag \\ 
		&\propto \exp \left\{ - \left( \vb{h} - \paramR{\mu}_{\vb{q}} \right)^\Hem \paramR{\Sigma}_{\vb{q}}^{-1} \left( \vb{h} - \paramR{\mu}_{\vb{q}} \right) \right\} \notag \\
			&\qquad \cdot \exp \left\{ - \left( \vb{h} - \paraml{\mu}_{\vb{q}} \right)^\Hem \paraml{\Sigma}_{\vb{q}}^{-1} \left( \vb{h} - \paraml{\mu}_{\vb{q}} \right) \right\}		
\end{align}
and the parameters of the distribution for $\nfg_{\vb{q} \rightarrow f_Y}(\vb{q})$ are found to be
\[ \paramr{\Sigma}_{\vb{q}} = \left( \paramR{\Sigma}_{\vb{q}}^{-1} + \paraml{\Sigma}_{\vb{q}}^{-1} \right)^{-1} \]
and
\[ \paramr{\mu}_{\vb{q}} = \paramr{\Sigma}_{\vb{q}} \left( \paramR{\Sigma}_{\vb{q}}^{-1} \paramR{\mu}_{\vb{q}} + \paraml{\Sigma}_{\vb{q}}^{-1} \paraml{\mu}_{\vb{q}}  \right) .\]

\section{Discrete BP for Equalization} \label{sec:equal_spa}
\subsubsection{Messages from auxiliary symbol variables to factor nodes}
The factors $f_{Y}$ receive posterior beliefs from the 
auxiliary variables as given by
\begin{equation}
	\nfg_{\vb{s} \rightarrow f_Y}\! (\vb{s}) \propto \mfg_{f_{Y} \rightarrow \vb{s}}(\vb{s}) \mfg_{f_{S} \rightarrow \vb{s}}(\vb{s}).
\end{equation}
On the other hand, messages passed from the auxiliary variables to the 
detection factors $f_S$ are in the form of extrinsic distributions
as given by
\begin{equation}
	\nfg_{\vb{s} \rightarrow f_{S}}\! (\vb{s}) \propto \mfg_{f_{Y} \rightarrow \vb{s}}(\vb{s}).
\end{equation}

\subsubsection{Messages from equalization node to auxiliary variables}
The BP rule (sum-product algorithm) 
leads to the following message:
\begin{align}
	\mfg_{f_{S} \rightarrow \vb{s}}(\vb{s}) & = \sum_{x_1} \cdots \sum_{x_N} p( \vb{s} | x_1, \ldots, x_N ) \prod_{i=1}^N \nfg_{x_i \rightarrow f_{S}}(x_i) \notag \\
		&= \prod_{i=1}^N \nfg_{x_i \rightarrow f_{S}}(s_i),
\end{align}
where $s_i$ is the $i$th element of $\vb{s}$ and
$p( \vb{s} | x_1, \ldots, x_N )$ enforces equality between
the symbol variables and the associated components of the auxiliary 
symbol variable as given by
\begin{equation}
	p( \vb{s} | x_1, \ldots, x_N ) = \prod_{i=1}^N \IndFn (s_i = x_i).
\end{equation}


\section{MF Message Derivations} \label{sec:mf_der}

Here we derive the messages involving the time-domain channel estimation in
the MF subgraph for the MIMO-OFDM model.

\subsubsection*{Observations to channel coefficients}
For the derivation of the message $\mfg_{f_{Y_k} \rightarrow h_{mn}(l)}(h_{mn}(l))$,
we first consider the factor function.
The factor function for factor node $f_{Y_k}$ is the likelihood
function of observation $\vb{y}(k)$.
Specifically we consider the likelihood function based on signal model 
\eqref{eq:signal_scalar} where the channel coefficients are expressed in terms 
of the time-domain taps as given by \eqref{eq:channel_individual}.
The likelihood function is given by 
\begin{align}
	&p(\vb{y}(k) | \vb{s}(k), \vb{h}) \notag \\ 
		&\quad = \left( \frac{\gamma}{\pi} \right)^M \exp \left\{ - \gamma \sum_{m=1}^{M} \left| y_m(k) - \sum_{n=1}^N s_n(k) \sum_{l=0}^{L-1} h_{mn}(l) d_{k l} \right|^2 \right\} \notag \\
		&\quad = \left( \frac{\gamma}{\pi} \right)^M \exp \left\{-\gamma \sum_{m=1}^{M} \left( \vphantom{\sum_{n=1}^N} |y_m(k)|^2 \right. \right.
				-2 \Re \left[ y_m(k)\sum_{n=1}^N s_{n}(k)^* \sum_{l=0}^{L-1} h_{mn}(l)^* d_{kl}^* \right] \notag \\
			&\qquad \quad + \sum_{n_1=1}^N \sum_{n_2=1}^N s_{n_1}(k) s_{n_2}(k)^* 
				\left. \left. \sum_{l_1=0}^{L-1} \sum_{l_2=0}^{L-1} h_{mn_1}(l_1)h_{mn_2}(l_2)^* d_{kl_1}d_{kl_2}^* \right) \right\}.
\end{align}
The factor function is simplified by removing all terms which are constant 
with respect to $h_{mn}(l)$ as given by
\begin{align}
	&p(\vb{y}(k) | \vb{s}(k), \vb{h}) \notag \\
	&\ \propto \exp \left\{ -\gamma \left( \vphantom{\sum_{n=1}^N} -2 \Re \left[ y_m(k) s_{n}(k)^* h_{mn}(l)^* d_{kl}^* \right]  + |s_n(k)|^2 |h_{mn}(l)|^2 \right. \right. \notag \\
		&\  \qquad + \left. \left. \! 2 \Re  \left[ \sum_{n^\prime \neq n} s_{n^\prime}(k) s_{n}(k)^* \sum_{l^\prime=0}^{L-1} h_{mn^\prime}(l^\prime) h_{mn}(l)^* d_{kl^\prime} d_{kl}^* + |s_{n}(k)|^2 \sum_{l^\prime \neq l} h_{mn}(l^\prime) h_{mn}(l)^* d_{kl^\prime} d_{kl}^* \vphantom{\sum_{n=1}^N} \right] \right) \right\} \notag \\
		&\ \propto \exp \left\{ -\gamma \left( |s_n(k)|^2 |h_{mn}(l)|^2 \notag \vphantom{\sum_n^N} \right. \right. 
			-2 \Re \left[ h_{mn}(l)^* \left( y_m(k) s_n(k)^* d_{kl}^* \vphantom{\sum_n^N} \right. \right. \notag \\
		&\ \qquad - \sum_{n^\prime \neq n} s_{n^\prime}(k) s_n(k)^* d_{kl}^* \sum_{l^\prime=0}^{L-1} h_{mn^\prime}(l^\prime) d_{kl^\prime} 
			\left. \left. \left. \left. - |s_n(k)|^2 d_{kl}^* \sum_{l^\prime \neq l} h_{mn}(l^\prime) d_{kl^\prime} \vphantom{\sum_n^N} \right) \right] \right) \right\}.
\end{align}

The message from observation factor node $f_{Y_k}$ to channel coefficient
$h_{mn}(l)$ is computed according to the MF approximation.
The derivation for the message is as follows:
\begin{align}
	&\mfg_{f_{Y_k} \rightarrow h_{mn}(l)}(h_{mn}(l)) \notag \\
		&\quad = \exp \Bigg\{ \idotsint \sum_{\vb{s}(k)} \nfg_{\vb{s}(k) \rightarrow f_{Y_k}}\!(\vb{s}(k)) \ln \left( p(\vb{y}(k) | \vb{s}(k), \vb{h}) \right)  \notag \\
			&\quad \qquad \smashoperator[r]{\prod_{\substack{ m^\prime,n^\prime,l^\prime \\ \{m^\prime,n^\prime,l^\prime\} \neq \{m,n,l\}}}} \nfg_{h_{m^\prime n^\prime}(l^\prime) \rightarrow f_{Y_k}}\!(h_{m^\prime n^\prime}(l^\prime)) dh_{m^\prime n^\prime}(l^\prime) \Bigg\} \notag \\
		&\quad \propto \exp \left\{ -\gamma \left( \paraml{\rho}_{\vb{s}(k)n,n} |h_{mn}(l)|^2 \vphantom{\sum_n^N} \right. \right.  -2 \Re \left[ h_{mn}(l)^* \left( y_m(k) \paraml{\mu}_{\vb{s}(k)n}^* d_{kl}^* \vphantom{\sum_n^N} \right. \right. \notag \\
			&\quad \qquad - \sum_{n^\prime \neq n} \paraml{\rho}_{\vb{s}(k)n^\prime,n} d_{kl}^*  \sum_{l^\prime=0}^{L-1} \mu_{h_{mn^\prime}(l^\prime)} d_{kl^\prime} 
			\left. \left. \left. \left. - \paraml{\rho}_{\vb{s}(k)n,n} d_{kl}^* \sum_{l^\prime \neq l} \mu_{h_{mn}(l^\prime)} d_{kl^\prime} \right) \right]
			\right) \right\}	\notag \\	
		&\propto \mathcal{CN}\left(h_{mn}(l) ; \, \phi_{mn}(l,k), \psi_{mn}(l,k)^{-1} \right),
\end{align}
where the mean is given by
\begin{equation}
	\phi_{mn}(l,k) = \paraml{\rho}_{\vb{s}(k)n,n}^{-1} \Bigg( y_m(k) \paraml{\mu}_{\vb{s}(k)n}^* d_{kl}^*
		- \sum_{n^\prime \neq n} \paraml{\rho}_{\vb{s}(k)n^\prime,n} d_{kl}^*  \sum_{l^\prime=0}^{L-1} \mu_{h_{mn^\prime}(l^\prime)} d_{kl^\prime}
		-\paraml{\rho}_{\vb{s}(k)n,n} d_{kl}^* \sum_{l^\prime \neq l} \mu_{h_{mn}(l^\prime)} d_{kl^\prime} \Bigg)
\end{equation}
and the precision (inverse variance) is given by
\begin{equation}
	\psi_{mn}(l,k) = \gamma \paraml{\rho}_{\vb{s}(k)n,n}.
\end{equation}
In the above expression, the mean $\paraml{\mu}_{\vb{s}(k)i}$ is the $i$th element of $\paraml{\mu}_{\vb{s}(k)}$ and cross-correlation $\paraml{\rho}_{\vb{s}(k)i,j}$ is the $i,j$th element of matrix $\paraml{\vb{R}}_{\vb{s}(k)}$.  


\subsubsection*{Channel coefficients to observations}
Since the observation factors are contained within the MF portion of the
graph, posterior beliefs are returned to them from the channel coefficient 
variables.  The message (posterior) is given by
\begin{align}
	&\nfg_{h_{mn}(l) \rightarrow f_{Y_k}}(h_{mn}(l)) \notag \\
		&\ = \prod_{k=0}^{K-1} \mfg_{f_{Y_k} \rightarrow h_{mn}(l)}(h_{mn}(l)) \mfg_{f_{h_{mn}(l)} \rightarrow h_{mn}(l)}(h_{mn}(l)) \notag \\
		&\ \propto \exp \left\{ - \left[ \sum_{k=0}^{K-1} \psi_{mn}(l,k) | h_{mn}(l) - \phi_{mn}(l,k) |^2
			+ \psi_{mn}^{p}(l) | h_{mn}(l) - \phi_{mn}^{p}(l) |^2 \vphantom{\sum_{k=0}^{K-1}} \right] \right\} \notag \\
		&\ \propto \exp \left\{ - | h_{mn}(l) |^2 \left( \sum_{k=0}^{K-1} \psi_{mn}(l,k) + \psi_{mn}^{p}(l) \right) \right. \notag \\
		&\ \ \left. + 2 \Re \left[ h_{mn}(l)\left( \sum_{k=0}^{K-1} \psi_{mn}(l,k) \phi_{mn}^*(l,k) + \psi_{mn}^{p}(l) \phi_{mn}^{p*}(l) \right) \right] \right\} \notag \\
		&\ \propto \mathcal{CN} \left( h_{mn}(l); \mu_{h_{mn}(l)} , \sigma_{h_{mn}(l)}^2 \right),
\end{align}
where the mean and variance of the message are given by 
\begin{equation}
	\mu_{h_{mn}(l)} = \frac{\sum_{k=0}^{K-1} \psi_{mn}(l,k) \phi_{mn}(l,k) + \psi_{mn}^{p}(l) \phi_{mn}^{p}(l)}{\sum_{k=0}^{K-1} \psi_{mn}(l,k) + \psi_{mn}^{p}(l)}
\end{equation}
and
\begin{equation}
	\sigma_{h_{mn}(l)}^2 = \left( \sum_{k=0}^{K-1} \psi_{mn}(l,k) + \psi_{mn}^{p}(l) \right)^{-1},
\end{equation}
respectively.  It is useful to denote the mean and variance of the 
frequency-domain channel coefficients per subcarrier.  According to
\ref{eq:channel_individual}, the mean and variance of $\tilde{h}_{mn}(k)$
are given by 
\begin{equation}
	\mu_{\tilde{h}_{mn}(k)} = \sum_{l=0}^{L-1} \mu_{h_{mn}(l)} d_{kl}
\end{equation}
and
\begin{equation}
	\sigma^2_{\tilde{h}_{mn}(k)} = \sum_{l=0}^{L-1} \sigma_{h_{mn}(l)}^2,
\end{equation}
respectively.

\subsubsection*{Observations to auxiliary variables}
The likelihood function based on signal model \eqref{eq:signal_vector}
is expressed as given by 
\begin{align}
	&p(\vb{y}(k) | \vb{s}(k), \tilde{\vb{H}}(k))  \notag \\
	&\quad = \exp \left\{ - \gamma \left( \vb{y}(k) - \tilde{\vb{H}}(k)\vb{s}(k) \right)^\Hem  \left( \vb{y}(k) - \tilde{\vb{H}}(k)\vb{s}(k)  \right)  \right\} \notag \\
	&\quad = \exp \left\{ - \gamma \left( \vb{y}(k)^\Hem \vb{y}(k) - 2\Re \left[ \vb{y}(k)^\Hem \tilde{\vb{H}}(k)\vb{s}(k) \right] 
		+ \vb{s}(k)^\Hem \tilde{\vb{H}}(k)^\Hem \tilde{\vb{H}}(k)\vb{s}(k)  \right)  \right\}.
\end{align}

In the constructed factor graph, the MF approximation message passing 
rule leads to the following message:
\begin{align}
	&\mfg_{f_{Y_k} \rightarrow \vb{s}(k)}(\vb{s}(k)) \notag \\
	&\quad = \exp \left\{ \idotsint \prod_{m,n,l} \nfg_{h_{mn}(l) \rightarrow f_{Y_k}}\! (h_{mn}(l)) 
		\ln p(\vb{y}(k) | \vb{s}(k), \tilde{\vb{H}}(k)) d\vb{h} \vphantom{\sum_n^N} \right\} \notag \\
	&\quad \propto \exp \left\{ -\gamma \left( \vb{s}(k)^\Hem \vb{W}(k) \vb{s}(k) - 2\Re \left[ \vb{y}(k)^\Hem \Xi(k) \vb{s}(k) \right] \right) \right\} \label{eq:m_obs_jsym_der} \\
	&\quad \propto \mathcal{CN} \left( \vb{s}(k); \vb{W}(k)^{-1} \Xi(k)^\Hem \vb{y}(k) , \gamma^{-1} \vb{W}(k)^{-1} \right), \label{eq:m_obs_jsym_pdf_der}
\end{align}
where 
\begin{align}
	\Xi(k) &= \idotsint \prod_{m,n,l} \nfg_{h_{mn}(l) \rightarrow f_{Y_k}}\! (h_{mn}(l)) \tilde{\vb{H}}(k) d\vb{h} \notag \\
	&= \begin{bmatrix}
		\mu_{\tilde{h}_{11}(k)} & \mu_{\tilde{h}_{12}(k)} & \cdots & \mu_{\tilde{h}_{1N}(k)} \\
		\mu_{\tilde{h}_{21}(k)} & \mu_{\tilde{h}_{22}(k)} & & \vdots \\
		\vdots & & \ddots & \\
		\mu_{\tilde{h}_{M1}(k)} & \cdots & & \mu_{\tilde{h}_{MN}(k)}
	\end{bmatrix}
\end{align}
and
\begin{align}
	&\vb{W}(k) = \idotsint \prod_{m,n,l} \nfg_{h_{mn}(l) \rightarrow f_{Y_k}}\! (h_{mn}(l)) \tilde{\vb{H}}(k)^\Hem \tilde{\vb{H}}(k) d\vb{h} \notag \\
		&\quad = \Xi(k)^\Hem \Xi(k) + \sum_{m=1}^M \text{diag} \left( \sigma^2_{\tilde{h}_{m1}(k)}, \sigma^2_{\tilde{h}_{m2}(k)}, \ldots, \sigma^2_{\tilde{h}_{mN}(k)} \right).
\end{align}
In the case in which $\vb{s}(k)$ is a discrete variable, the message may be determined 
(up to a multiplicative constant)
by evaluating \eqref{eq:m_obs_jsym_der} with respect to $\vb{s}(k)$
which avoids the matrix inversion of \eqref{eq:m_obs_jsym_pdf_der}.

\bibliographystyle{IEEEtran}
\bibliography{IEEEabrv,master}

\end{document}